\documentclass[australian,english,prl, singlespace, twocolumn]{revtex4-1}
\usepackage[T1]{fontenc}
\usepackage[latin9]{inputenc}
\usepackage{color}
\usepackage{amsmath}
\usepackage{amssymb}
\usepackage{graphicx}

\makeatletter
\@ifundefined{definecolor}
 {\usepackage{color}}{}
\makeatother

\makeatother

\usepackage{babel}
\begin{document}
\title{An Einstein-Podolsky-Rosen argument based on weak forms of local realism
not falsifiable by GHZ or Bell experiments}
\author{Jesse Fulton, Run Yan Teh and M. D. Reid$^{1}$}
\affiliation{$^{1}$ Centre for Quantum Science and Technology Theory, Swinburne
University of Technology, Melbourne 3122, Australia}
\begin{abstract}
The Einstein-Podolsky-Rosen (EPR) paradox gives an argument for
the incompleteness of quantum mechanics based on the premises of local
realism. A general view is that the argument is compromised, because
EPR's premises are falsified by Greenberger-Horne-Zeilinger (GHZ)
and Bell experiments. In this paper, we present an EPR argument based
on premises not falsifiable by these experiments. First, we propose
macroscopic EPR and GHZ experiments using  spins $\hat{S}_{\theta}$
defined by two macroscopically distinct states. The analyzers that
realize the unitary operations $U_{\theta}$ determining the measurement
settings $\theta$ are nonlinear devices known to create macroscopic
superposition states. We note two definitions of macroscopic realism
(MR). For a system with two macroscopically distinct states available
to it, MR posits a predetermined outcome for a measurement $\hat{S}_{\theta}$
distinguishing between the states. Deterministic macroscopic realism
(dMR) posits MR for the system defined\emph{ prior} to the interaction
$U_{\theta}$ being carried out.  Weak macroscopic realism (wMR)
posits MR for the system  \emph{after} $U_{\theta}$, at the time
$t_{f}$ $-$ when the system is prepared with respect to the measurement
basis, ready for a final ``pointer'' measurement and readout.
For this system, wMR posits that the outcome of $\hat{S}_{\theta}$
is determined, and not changed by interactions that might occur at
a remote system $B$. The premise wMR also posits that if the outcome
for $\hat{S}_{\theta}^{A}$ of a system $A$ can be predicted with
certainty by a ``pointer'' measurement on the system $B$ defined
at time $t_{f}^{B}$ \emph{after} the unitary interaction that fixes
the setting at $B$, then the outcome for $\hat{S}_{\theta}^{A}$
is determined for $A$ at this time $t_{f}^{B}$, regardless of whether
the unitary interaction required to fix the setting as $\theta$ has
taken place at $A$. We show that the GHZ predictions negate dMR
but are consistent with wMR. Yet, an EPR paradox  arises based on
wMR. As considered by Schrödinger, it is possible to measure two complementary
spins of system $A$ simultaneously, ``one by direct, the other by
indirect measurement'': If we assume wMR, then at time $t_{f}=t_{f}^{B}$,
the outcomes of the two spins are both  determined.\emph{ }We
revisit the original EPR paradox and find a similar result: An EPR
argument can be based on a weak contextual form of local realism (wLR)
not falsifiable by GHZ or Bell tests.

\end{abstract}
\maketitle

\section{Introduction}

In their argument of 1935, Einstein-Podolsky-Rosen (EPR) introduced
premises, based on local realism, which if valid suggested quantum
mechanics to be an incomplete description of physical reality \cite{epr}.
The argument considered two separated particles with correlated positions
and anticorrelated momenta. The correlations imply that the outcome
of a measurement of either position or momentum could be inferred
with certainty for one particle, by an experimentalist making the
appropriate measurement on the second particle. Assuming there is
no disturbance to the first particle by the experimentalist's actions,
EPR argued from their premises that the position and momentum of the
first particle are both simultaneously precisely determined prior
to measurement, thereby creating an inconsistency with any quantum-state
description for the localised particle. 

Bell later proved that all local realistic theories could be falsified
by quantum predictions \cite{bell-1969,bell-cs-review,bell-1971,chsh,bell-brunner-rmp}.
Moreover, Greenberger-Horne-Zeilinger (GHZ) \cite{ghz-1,mermin-ghz,ghz-amjp,clifton-ghz}
gave a falsification of EPR's premises in an ``all or nothing''
situation.  Bell and GHZ predictions have been experimentally verified
\cite{ghz-pan,ghz-exp-Bou}. Consequently, the EPR paradox is most
often regarded as an illustration of the incompatibility between local
realism and quantum mechanics, rather than as a valid argument for
the incompleteness of quantum mechanics \cite{epr-rmp,mermin-ghz}.

In this paper, we present a different perspective on the EPR paradox.
We first propose a test of local realism versus quantum mechanics
in a macroscopic GHZ set-up, where realism refers to a system being
in one or other of two macroscopically distinct states. This provides
a way to falsify local realism at a macroscopic level. Given such
a falsification may raise questions for the interpretation of quantum
measurement \cite{legggarg-1,s-cat-1935}, we then examine carefully
the definitions of macroscopic realism, showing that a less restrictive
definition of macroscopic realism is not falsified by GHZ or Bell
experiments. This leads to a second conclusion: A modified EPR argument
that quantum mechanics is incomplete can be given, based on an alternative
and (arguably) nonfalsifiable premise.

Specifically, we show how the GHZ and EPR paradoxes can be realised
for macroscopic qubits, where all relevant measurements $\hat{S}_{\theta}$
distinguish between two macroscopically distinct states. The EPR paradox
is presented as Bohm's version \cite{Bohm,bohm-aharonov,aspect-Bohm}
which examines two spatially-separated entangled spin-$1/2$ systems.
The macroscopic version is a direct mapping of the original paradox,
where a spin $|\uparrow\rangle$ and $|\downarrow\rangle$ is realised
as two macroscopically distinct states, such as coherent states $|\alpha\rangle$
and $|-\alpha\rangle$, or collective multimode spin states $\prod_{k=1}^{N}|\uparrow\rangle_{k}$
and $\prod_{k=1}^{N}|\downarrow\rangle_{k}$. The necessary unitary
transformations $U_{\theta_{i}}$ which determine the measurements
settings $\theta_{i}$ for each system $i$ are realised  by nonlinear
interactions, or  CNOT gates.

Leggett and Garg gave a definition of macroscopic realism (MR) for
a system ``with two or more macroscopically distinct states available
to it'': MR asserts that the system ``will at all times be in one
or other of those states'' \cite{legggarg-1}.  Following previous
work, we note different definitions are possible \cite{manushan-bell-cat-lg,delayed-choice-cats}.
\emph{Deterministic macroscopic (local) realism} (dMR) asserts there
is a predetermined value $\lambda$ for the outcome of a measurement
$S$ that will distinguish between the macroscopically distinct states.
Locality is implied, since it is assumed that this value is not affected
by spacelike-separated interactions or events.

However, the EPR-Bohm, Bell and GHZ experiments require choices of
measurement settings $\theta_{i}$ at each site $i$, the choice establishing
which spin component $S_{\theta_{i}}$ will be measured. This leads
to different definitions of MR. The measurement basis is determined
by the setting of a physical device (analogous to a Stern-Gerlach
analyzer) which realizes a unitary operation $U_{\theta_{i}}=e^{-iH_{\theta_{i}}t/\hbar}$
where $H_{\theta_{i}}$ is the interaction Hamiltonian.  After
the interaction $U_{\theta_{i}}$ at a site $i$, there is a final
stage of measurement that includes an irreversible coupling to an
environment to give a readout of the spin $S_{\theta_{i}}$. We refer
to this final stage as the ``pointer measurement''.

In the macroscopic experiments we propose, the system after the selected
interactions $U_{\theta_{i}}$ is in a superposition of macroscopically
distinct states which have definite values for the outcomes $S_{\theta_{i}}$.
\emph{Weak macroscopic realism} (wMR) posits that each system $i$
 prepared at a time $t_{f_{i}}$ \emph{after} the interaction $U_{\theta_{i}}$
can be ascribed a predetermined value $\lambda_{\theta_{i}}$ for
the final pointer part of the measurement $\hat{S}_{\theta_{i}}$,
which will distinguish between the macroscopic states. 

The premise wMR posits not only a weak form of realism, but also a
weak form of locality. Consider the local system $i$ prepared at
the time $t_{f_{i}}$ for the pointer measurement: There is no disturbance
to the value $\lambda_{\theta_{i}}$ for the pointer measurement from
interactions that might occur at a spacelike-separated site; nor \emph{from}
the pointer measurement (if it happened to be carried out) to a spacelike-separated
system. In such a model, we will show that the nonlocal effects
contributing to the GHZ and Bell contradictions with local realism
emerge when there are further unitary interactions occurring at \emph{both}
sites.

The premise of deterministic local macroscopic realism (dMR) is similar
to EPR and Bell's form of local realism, because the premise applies
to the system as it exists \emph{prior} to the unitary interactions
$U_{\theta_{i}}$. The macroscopic GHZ set-up hence enables an ``all
or nothing'' falsification of dMR, which supports previous work revealing
dMR to be falsifiable by macroscopic Bell tests \cite{macro-bell-lg,manushan-bell-cat-lg,delayed-choice-cats,macro-bell-jeong,wigner-friend-macro,bell-contextual}.

The main result of this paper is that weak macroscopic realism (wMR)
is \emph{not} falsified by the GHZ or Bell set-ups. Yet, an EPR-type
argument can be put forward based on wMR. The argument applies to
the EPR setup considered by Schrödinger, where two noncommuting observables
are measured simultaneously ``\emph{one by a direct, the other by
indirect measurement}'' \cite{sch-epr-exp-atom,s-cat-1935}. In our
paper, the EPR-Bohm gedanken experiment is examined at the time $t_{f}$
after the unitary interactions $U_{\theta_{i}}$ have been carried
out at each site, in order to measure spin $S_{z}$ of one system
and $S_{y}$ of the other. The premise wMR posits a predetermined
value for the pointer measurement of $S_{z}$ of the first system,
since it can be shown to have two ``macroscopically distinct states
available to it'' (with respect to the measurement basis $S_{z}$).
But also for the EPR-Bohm state, the outcome for $S_{y}$ can be inferred
for the first system by a pointer measurement on the other. Hence,
wMR posits \emph{simultaneous precise values for both $S_{z}$ and
$S_{y}$}. This is not consistent with a local quantum state, the
Pauli spin variances being constrained by $(\Delta\sigma_{z})^{2}+(\Delta\sigma_{y})^{2}\geq1$
\cite{hofmann-take}, which leads to the paradox. It is important
to note that there is no actual violation of the uncertainty principle,
because the quantum state defined at the time $t_{f}$ is different
to the quantum state defined after the further interaction $U_{\theta'_{i}}^{A}$
necessary to change the measurement setting from $z$ to $y$ at $A$.

The results motivate us to revisit the original microscopic EPR-Bohm
paradox, and to demonstrate an EPR argument based on a \emph{weak
contextual form of local realism} (wLR), which we show is not falsified
by GHZ or Bell set-ups. The definitions of wMR and wLR are both
contextual, being defined for the system with a specified measurement
basis. The weaker assumptions are motivated by Bohr's criticism
of EPR's 1935 paper \cite{bell-cs-review,bohr-epr}. Clauser and
Shimony state that ``\emph{{[}Bohr's{]} argument is that when the
phrase \textquoteleft without in any way disturbing the system\textquoteright{}
is properly understood, it is incorrect to say that system 2 is not
disturbed by the experimentalist\textquoteright s option to measure
$a$ rather than $a'$ on system 1.}'' This suggests that \emph{after}
the experimentalist's option to measure say the spin component $a$,
there is reason to justify no disturbance, the nonlocality stemming
from the unitary interactions giving the options. We explain
how wLR may be implied by wMR, by considering the system at the time
of the reversible coupling to a macroscopic meter.

The layout of the paper is as follows. In Section II, we outline the
definitions of local realism and macroscopic realism used in this
paper. In Section III, we review the original EPR-Bohm and GHZ paradoxes,
giving in Section IV the modified EPR-Bohm paradox based on the premise
of wLR. In Sections V and VII, we present the proposals for macroscopic
EPR-Bohm and GHZ paradoxes using cat states. Conclusions drawn from
these paradoxes are given in Sections VI and VIII. The macroscopic
GHZ paradox falsifies dMR, but shows consistency with wMR. Similarly,
the GHZ paradox is consistent with wLR. In Section IX, we demonstrate
consistency of wLR (and wMR) with violations of Bell inequalities.
Further tests of wMR and wLR are devised in Section X.

\section{Definitions}

We formalize the definitions of local realism and macroscopic realism
relevant to this paper. Several different definitions are introduced.
The difference between the definitions is clarified, once we recognise
that there are two stages to a spin measurement $S_{\theta}$: First,
there is the reversible measurement-setting stage involving a unitary
interaction $U_{\theta}$ which determines the measurement setting
$\theta$. Second, there is the stage that comes after, which includes
a final irreversible readout of a meter. We refer to the later stage
as the \emph{pointer {[}stage of{]} measurement.}

Consider two separated spin-$1/2$ systems $A$ and $B$ prepared
at time $t_{0}$ in the state $|\psi\rangle$. Local unitary interactions
$U_{\theta}^{A}$ and $U_{\phi}^{B}$ prepare the systems for spin
measurements $S_{\theta}^{A}$ and $S_{\phi}^{B}$. The $U_{\theta}$
are realised as reversible interactions of the system with a real
device, such as a Stern-Gerlach analyzer or polarizing beam splitter,
and are represented by a Hamiltonian $H_{\theta}$, where $U_{\theta}=e^{-iH_{\theta}t/\hbar}$.
The $U_{\theta}$ takes place over a time interval $t$, and the states
prior and after the interaction $U_{\theta}$ may therefore be regarded
as different, in that they define the system at a different time.
The state after the interaction at time $t_{f}$ is 
\begin{equation}
|\psi(t_{f})\rangle=e^{-iH_{\theta}t_{f}/\hbar}|\psi\rangle\label{eq:state-after-1}
\end{equation}
Specific examples are given in Sections IV and V, where we note that
the ``system'' may include another (local) set of modes, or a meter
that is originally decoupled to the spin system, in which case $|\psi\rangle$
is suitably defined. After the interaction $U$, the final irreversible
stage of the measurement is made, which indicates the measurement
outcome. This stage often involves a direct detection of a particle
at a given location, as well as amplification and a coupling to a
meter. The local system prepared after the interaction $U$ that
fixes the measurement setting, but before the irreversible stage of
the measurement, is considered to be \emph{prepared for the pointer
measurement}. The system is said to be prepared in the \emph{preferred
basis}, also referred to as the \emph{measurement, or pointer, basis}.

A common realisation is the spin $1/2$ system $|\uparrow\rangle\equiv|1,0\rangle$
and $|\downarrow\rangle\equiv|0,1\rangle$ defined for two orthogonally
polarized modes $a_{\pm}$. Here, $|n_{1},n_{2}\rangle\equiv|n_{1}\rangle_{+}|n_{2}\rangle_{-}$
where $|n\rangle_{\pm}$ is a number state for the mode $a_{\pm}$.
A transformation $U_{\theta}$ can then be achieved with a polarizing
beam splitter, with mode transformations ($\hat{a}_{\pm}$ are boson
operators defining the modes)
\begin{eqnarray}
\hat{c}_{+} & = & \hat{a}_{+}\cos\theta-\hat{a}_{-}\sin\theta\nonumber \\
\hat{c}_{-} & = & \hat{a}_{+}\sin\theta+\hat{a}_{-}\cos\theta.\label{eq:bs-1}
\end{eqnarray}
The $\hat{c}_{\pm}$ are boson operators for the outgoing modes emerging
from the beam splitter. The interaction is described by the Hamiltonian
$H_{\theta}=i\hbar k(\hat{a}_{+}\hat{a}_{-}^{\dagger}-\hat{a}_{+}^{\dagger}\hat{a}_{-})$
where $\theta=kt$. The choice $\theta=0$ ($\theta=\pi/4)$ corresponds
to a measurement of $\sigma_{z}$ ($\sigma_{x}$). A single photon
impinges on the beam splitter and is finally detected at one or other
locations associated with the outgoing modes \cite{aspect-Bohm}.
The final detection and readout of the locations constitutes the pointer
measurement.

More recently, an EPR experiment has been realised for pseudo-spin
measurements in a BEC setting \cite{sch-epr-exp-atom}. Here, the
measurement setting is determined by an interaction of a field with
a two-level atom, which forms the spin $1/2$ system.
\begin{figure}[t]
\begin{centering}
\includegraphics[width=1\columnwidth]{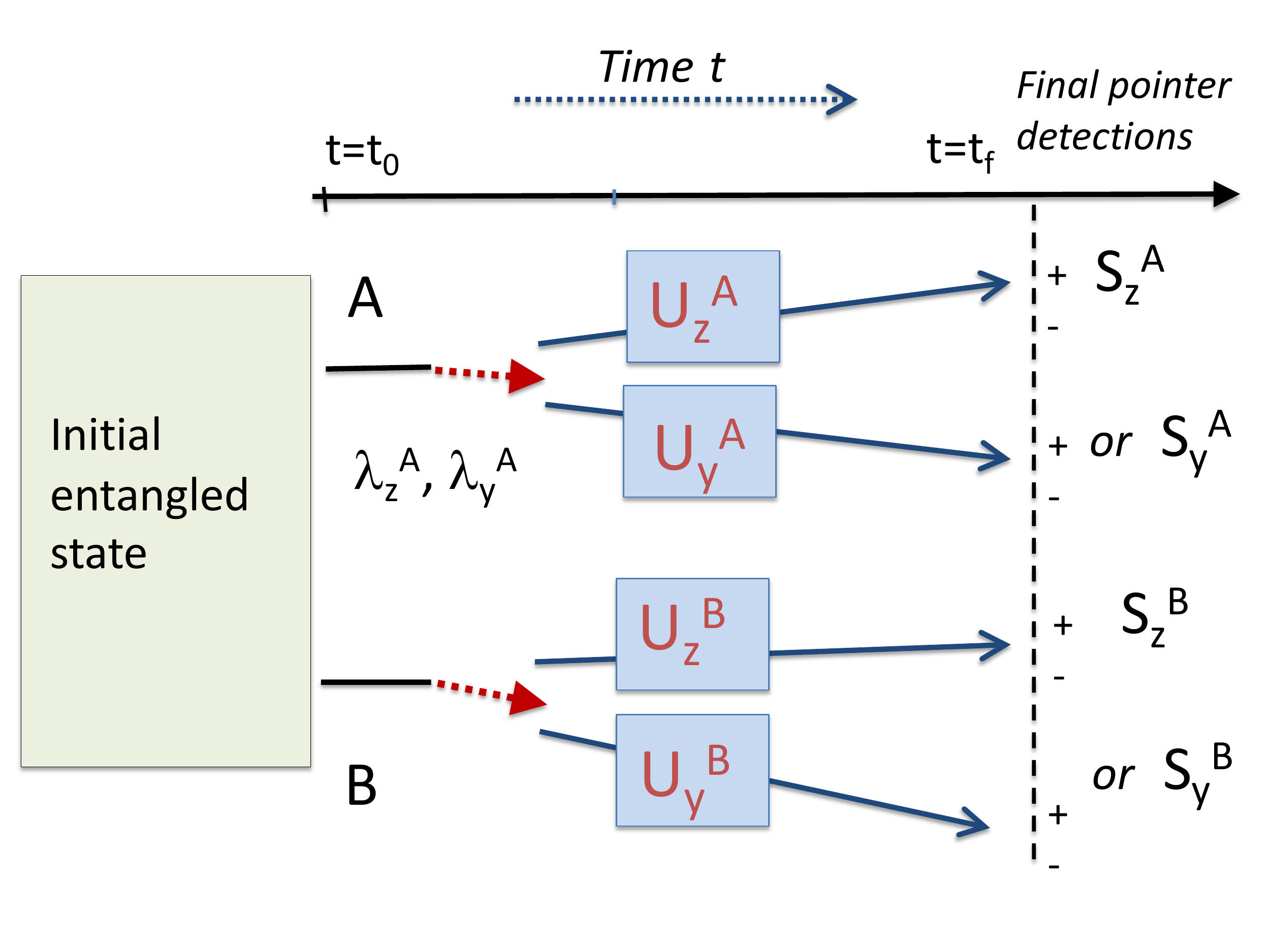}
\par\end{centering}
\caption{The set-up for an EPR-Bohm paradox. Two separated systems $A$ and
$B$ are prepared in an entangled state $|\psi\rangle$ (Eq. (\ref{eq:bell})).
A switch (red dashed arrow) gives the choice to measure either $S_{z}$
or $S_{y}$ for each of $A$ and $B$, by passing through an appropriately
orientated analyzer symbolised by $U$ at each site. Here, $S_{z}$
and $S_{y}$ correspond to Pauli spins $\sigma_{z}$ and $\sigma_{y}$.
EPR's local realism asserts that because one can predict the result
for $S_{z}$ (or $S_{y}$) at $A$ by measuring $S_{z}$ (or $S_{y}$)
at $B$, the outcomes for $S_{z}$ and $S_{y}$ at $A$ are \emph{both}
predetermined, at the time $t_{0}$ (prior to the choices of measurement
setting). The premises of macroscopic local realism and deterministic
macroscopic realism apply when the outcomes $+$ and $-$ for both
spins $S_{z}$ and $S_{y}$ are associated with macroscopically distinct
states for the system at the time $t_{0}$. \label{fig:Sketch-bohm-test}}
\end{figure}

\subsection{Strong elements of reality: EPR's Local realism and Macroscopic realism}

\subsubsection{EPR's Local realism}

The premises presented in the original 1935 argument given by EPR
are based on the assumptions of local realism. The premises, which
we refer to as EPR's local realism (LR), are summarized as two Assertions
 for space-like separated systems, $A$ and $B$.

\paragraph*{EPR's Assertion LR (1): Realism}

EPR's reality criterion is: ``If, without in any way disturbing a
system, we can predict with certainty the value of a physical quantity,
then there exists an element of physical reality corresponding to
this physical quantity \cite{epr}.''

This is interpreted as follows: ``The \textquotedbl element of
physical reality\textquotedbl{} is that predictable value, and it
ought to exist whether or not we actually carry out the procedure
necessary for its prediction, since that procedure in no way disturbs
it \cite{mermin-ghz}.'' Hence, EPR Assertion LR (1) reads: If one
can predict with certainty the outcome of a measurement $S$ on system
$A$ without disturbing that system, then the outcome of the measurement
is predetermined. The system $A$ can be ascribed a variable $\lambda^{A}$,
the value of which gives the outcome for $S$ \cite{mermin-ghz}.

\paragraph*{EPR Assertion LR (2): No disturbance (Locality)}

There is no disturbance to system $A$ from a spacelike-separated
interaction or event (e.g. a measurement on system $B$).

The consequences of the two Assertions as applied to the EPR-Bohm
set-up leads to the EPR-Bohm paradox (Section III). Consider the system
of Figure \ref{fig:Sketch-bohm-test}: If the outcome of the measurement
$S_{\theta}^{A}$ at $A$ can be predicted with certainty by a measurement
at $B$, then the EPR Assertions imply the system $A$ at time $t_{0}$
can be ascribed a variable $\lambda_{\theta}^{A}$, the value of which
gives the outcome of the measurement $S_{\theta}^{A}$. The assignment
of the variable $\lambda_{\theta}^{A}$ can be made to the system
$A$ regardless of the measurement device actually being prepared,
either at $A$ or at $B$. In the Figure \ref{fig:Sketch-bohm-test},
this allows the assignment of \emph{both} variables $\lambda_{z}^{A}$
and $\lambda_{y}^{A}$ to system $A$, at the time $t_{0}$, prior
to the unitary interactions $U_{\theta}^{A}$ and $U_{\phi}^{B}$
that determine the measurement settings. 

\subsubsection{Macroscopic local realism and deterministic macroscopic realism}

The assertions defining \emph{macroscopic local realism} (MLR) are
identical to those of EPR's local realism, except that the assertions
are weaker, being restricted to apply only to the subset of systems
where the outcomes for all relevant measurements, $S_{\theta}^{A}$
and $S_{\phi}^{B}$, correspond to macroscopically distinct states
of the system. This means that the systems upon which those measurements
are made can be viewed as having two (or more) macroscopically distinct
states available to them, so that the Leggett-Garg definition of \emph{macroscopic
realism} \cite{legggarg-1} can be applied, to separately posit \emph{deterministic
macroscopic realism} (dMR), as below. It would be argued that the
assumption of realism is more robustly justified for macroscopically
distinct states \cite{s-cat-1935}.

\paragraph*{Assertion MLR (1a): EPR's realism}

This reads as for EPR Assertion LR (1). ``If, without in any way
disturbing a system, we can predict with certainty the value of a
physical quantity, then there exists an element of physical reality
corresponding to this physical quantity.''

\paragraph*{Assertion dMR (1b): Leggett-Garg's criterion for Macroscopic realism\emph{
}$-$ Deterministic macroscopic realism (dMR)}

A system which has two or more macroscopically distinct states available
to it can be ascribed a predetermined value $\lambda$ for the measurement
$S$ that will distinguish between these states. The predetermined
value $\lambda$ is not affected by spacelike-separated measurements
(e.g. further unitary transformations $U_{\phi}$) that may occur
at site $B$. Hence, locality is implied, which also follows from
Assertion MLR(2) below.

\paragraph*{Assertion MLR (2): No disturbance (Locality)}

There is no macroscopic disturbance to system $A$ from a spacelike-separated
interaction or event.

Similar to EPR's local realism, deterministic macroscopic realism
asserts there is a predetermined value $\lambda_{\theta}$ for the
outcome of the measurement at the time $t_{0}$, for the system as
it exists prior to, or irrespective of, the interaction $U_{\theta}$
(Figure \ref{fig:Sketch-bohm-test}).

\subsection{Weak elements of reality: weak macroscopic realism and weak local
realism}

\subsubsection{Weak macroscopic realism}

\begin{figure}[t]
\begin{centering}
\includegraphics[width=1\columnwidth]{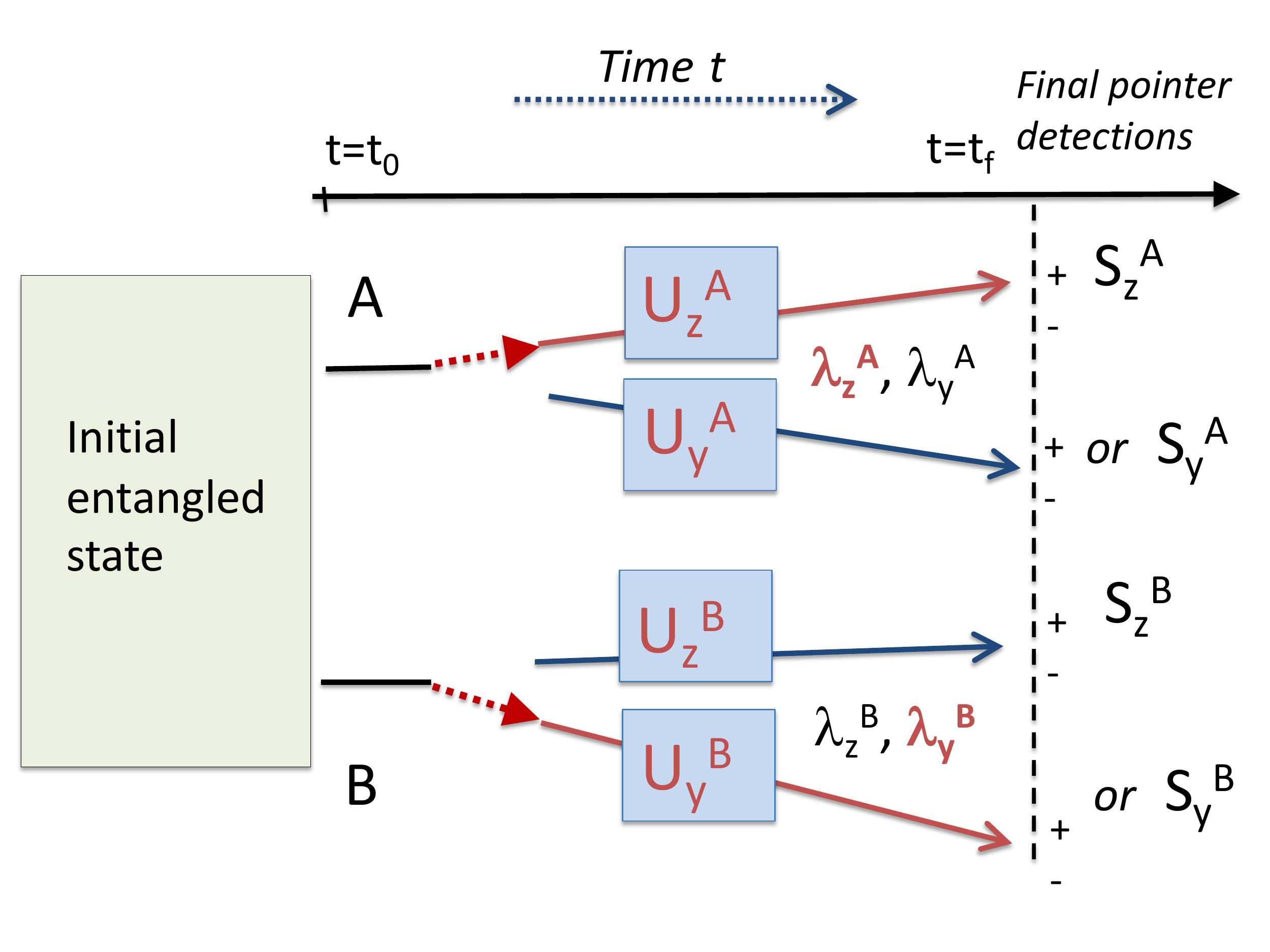}
\par\end{centering}
\caption{The assumption of weak local realism (wLR) gives rise to an EPR paradox:
The set-up is for an EPR-Bohm paradox as in Figure \ref{fig:Sketch-bohm-test}.\emph{
}At time $t_{f}$, the measurement settings are set to $S_{z}^{A}$
and $S_{y}^{B}$, as indicated by the positions of the red dashed
arrows. Weak local realism asserts that for the system $A$ at time
$t_{f}$, after the unitary rotation $U_{z}^{A}$, the outcome for
the final pointer measurement $S_{z}$ at $A$ is determined, given
by the variable $\lambda_{z}^{A}$. Weak local realism also asserts
that because one can predict the result for $S_{y}$ at $A$ by making
a final pointer measurement $S_{y}$ at $B$, the outcome for $S_{y}$
at $A$ is also determined at time $t_{f}$ (after the rotation $U_{y}^{B}$
at $B$). This is despite that a further unitary evolution $U^{A}$
would be required at $A$ to perform the measurement $S_{y}^{A}$.
This leads to the paradox. The premise of weak macroscopic realism
applies similarly, assuming the outcomes $+$ and $-$ for the spins
can be viewed as corresponding to macroscopically distinct states
for the system at the time $t_{f}$. \label{fig:Sketch-bohm-test-2}}
\end{figure}

Weak macroscopic realism (wMR) involves weaker (i.e. less restrictive)
assumptions than macroscopic local realism. Macroscopic local realism
implies wMR, but the converse is not true. The assertions for wMR
are modified so that EPR's local realism applies to the systems \emph{after}
the selection of the measurement settings, at time $t_{f}$ in the
Figure \ref{fig:Sketch-bohm-test-2}.

\paragraph*{Assertion wMR (1a): EPR's criterion for realism}

This assertion reads as for Assertion MLR (1a). ``If, without in
any way disturbing a system, we can predict with certainty the value
of a physical quantity, then there exists an element of physical reality
corresponding to this physical quantity \cite{epr}.''

The conclusions from Assertion wMR (1a) will now be different from
those of EPR, due to the modification of Assertion (2a) below. Suppose
system $B$ at time $t_{f}$, after the unitary interaction $U_{\phi}^{B}$,
 is prepared for the pointer stage of measurement of spin $S_{\phi}^{B}$.
Assertion wMR 2(a) asserts that there is no disturbance to system
$A$ due to whether or not the pointer stage of measurement at $B$
\emph{actually} takes place. The modification means that the prediction
for the outcome of measurement $S$ at $A$, as based on a measurement
at $B$, ensures a predetermination of the result at $A$ at the time
$t_{f}$, provided the unitary interaction $U_{\phi}^{B}$ that fixes
the measurement setting $\phi$ at $B$ has taken place.

The Assertion wMR (1a) can hence be rephrased: Consider the system
of Figure \ref{fig:Sketch-bohm-test-2}: If the outcome of a measurement
$S_{\theta}^{A}$ at $A$ can be predicted with certainty by a \emph{pointer}
measurement on the system at time $t_{f}$ at $B$, then there exists
an element of physical reality corresponding to the outcome of $S_{\theta}^{A}$
at $A$. Thus, the system $A$ can be ascribed a hidden variable $\lambda_{\theta}^{A}$
that determines the outcome for $S_{\theta}^{A}$. This is true regardless
of whether the pointer measurement at $B$ is actually carried out
(because that would not disturb the system $A$), and regardless of
whether the interaction $U^{A}$ at $A$ that fixes the measurement
setting $\theta$ has actually been carried out at $A$ (and regardless
of future unitary interactions at $A$). However, the predetermination
is based on the system $B$ being prepared for a pointer measurement,
and therefore only applies at the time $t_{f}$, when no further unitary
interactions that would cause a change of measurement setting at $B$
have taken place.

\paragraph*{Assertion wMR (1b): Macroscopic realism for the system prepared for
the pointer measurement}

Suppose a system $A$ that is prepared for the pointer measurement
of $S$ can be considered to have two or more macroscopically distinct
states available to it, where each of those states has a definite
outcome for the pointer measurement. Weak macroscopic realism asserts
that such a system can be ascribed a predetermined value $\lambda^{A}$
for the pointer measurement $S$ that will distinguish between these
states.

The premise applies, so that the result of the pointer measurement
for $S_{\theta}^{A}$ is predetermined at the time $t_{f}$, once
the interaction $U_{\theta}^{A}$ determining the measurement setting
at $A$ has taken place (Figure \ref{fig:Sketch-bohm-test-2}). 

\paragraph*{Assertion wMR (2a): Pointer locality: No disturbance from a pointer
measurement}

The pointer stage of a measurement gives no disturbance to a spacelike-separated
system i.e. there is no disturbance to system $A$ from a pointer
measurement on system $B$.

\paragraph*{Assertion wMR (2b): Pointer locality: No disturbance to a pointer
measurement}

There is no disturbance to the predetermined value $\lambda^{A}$
for the pointer measurement at $A$ (as described in Assertion wMR
(1b)) by spacelike-separated interactions or events (e.g. further
unitary transformations $U_{\phi}^{B}$) that may occur at $B$.

The assertions when applied to EPR-Bohm set-up of Figure \ref{fig:Sketch-bohm-test-2}
will lead to an EPR-type paradox, as explained in Sections IV and
V.

\subsection{Weak local realism}

The premise of \emph{weak local realism} (wLR) is a weak form of
local realism, which applies to the system prepared at time $t_{f}$
for the pointer measurement (Figure \ref{fig:Sketch-bohm-test-2}).
This contrasts with definitions of local realism that apply to the
system at time $t_{0}$, \emph{prior} to the entire measurement process,
and which can be falsified. It should be mentioned that weak local
realism, as with weak macroscopic realism, does \emph{not }exclude
``nonlocality'', since, as we will see, these premises are consistent
with the quantum predictions for Bell and GHZ experiments.

The assertions of wLR are as for weak macroscopic realism (wMR), except
there is no longer the restriction that the outcomes correspond to
macroscopically distinct states of the system being measured. A
connection between wLR and wMR is given in the Section II.D. As with
wMR, in  this model, the pointer measurement constitutes a passive
stage of the measurement.

We comment on the terminology\emph{ local realism}. The most general
definition of local realism is given by Bell's local hidden variable
theories, also referred to as local realistic theories \cite{bell-cs-review}.
These theories allow for local interactions with local measurement
devices, so that the value for the outcome of the measurement need
not be predetermined (at time $t_{0}$ or $t_{f}$). This contrasts
with the stricter definition, which we refer to as a non-contextual
or deterministic local realism \cite{bell-cs-review}, meaning the
values for measurement outcomes are predetermined at the time $t_{0}$,
prior to the entire measurement process including the unitary interactions
$U$. However, general local realistic hidden variable theories \emph{imply}
EPR's local realism. Moreover, for the EPR-Bohm system where the correlations
between certain spin measurements are maximum, the local realistic
theories also imply the stricter deterministic local realism, for
spin measurements \cite{bell-1969}.

Generally speaking, however, we cannot assume wLR to be a ``weaker''
assumption than local realism, in the sense that it not necessarily
a subset of those assumptions, since local realistic theories may
allow non-passive interactions with the pointer measurement apparatus.
To avoid confusion, we emphasize that weak local realism refers to
a weaker version of the \emph{non-contextual deterministic} form of
local realism. In later Sections, we show that while \emph{all} local
realistic theories are ruled out by GHZ and Bell experiments, wLR
theories are not, which motivates the terminology ``weak''.

\paragraph*{Assertion wLR (1a): EPR's realism}

The assertion is as for EPR Assertion LR (1). As with wMR, the conclusions
drawn from this Assertion are impacted by Assertion (2a). Weak
local realism asserts: If the outcome of the measurement $S_{\theta}^{A}$
at $A$ can be predicted with certainty at time $t_{f}$ by the final
pointer stage of measurement at $B$ (the local dynamics $U$ of the
measurement setting being already performed at $B$), then the system
at $A$ at time $t_{f}$ can be ascribed a variable $\lambda_{\theta}^{A}$,
the value of which gives the outcome of the measurement $S_{\theta}^{A}$.
This is true regardless of whether the pointer measurement at $B$
is actually performed, and regardless of whether the unitary interaction
$U_{\theta}^{A}$ has taken place at $A$, and regardless of any further
unitary interactions at $A$ (Figure \ref{fig:Sketch-bohm-test-2}).

\paragraph*{Assertion wLR (1b): Realism for the system prepared for the pointer
measurement}

The result of the pointer measurement for $S^{A}$ for system $A$
is predetermined (by a variable $\lambda^{A}$) once the local interaction
$U_{\theta}^{A}$ for the measurement setting at $A$ has taken place.

Consider the system $A$ at time $t_{f}$, after the unitary rotation
$U_{\theta}^{A}$. The system $A$ is prepared for the pointer stage
of measurement of $S_{\theta}^{A}$, without the need for a further
unitary rotation $U$. The premise wLR asserts that the system $A$
at time $t_{f}$ can be ascribed by a hidden variable $\lambda_{\theta}^{A}$
with value $+1$ or $-1$, that value determining the outcome of the
pointer measurement for $S_{\theta}^{A}$ at $A$, if that pointer
stage of measurement were to be carried out on the prepared state.
In accordance with Assertion (2b), the predetermined value $\lambda_{\theta}^{A}$
is not affected by spacelike-separated events (e.g. further unitary
transformations $U_{\phi}$) that may occur at $B$.

\paragraph*{Assertion wLR (2a): Pointer locality: No disturbance from a pointer
measurement}

The assertion reads as for Assertion wMR (2a).

\paragraph*{Assertion wLR (2b): Pointer locality: No disturbance to a pointer
measurement}

The assertion reads as for Assertion wMR (2b).

The Assertion wLR (1b) which asserts realism is less convincing for
a microscopic system than for a macroscopic system, and might seem
to contradict the results of Bell's theorem. We later show that wLR
is not contradicted by the Bell or GHZ predictions. The assertions
when applied to the set-up of Figure \ref{fig:Sketch-bohm-test-2}
will nonetheless lead to an EPR-type paradox, as explained in Section
IV.

\subsection{Link between wLR and wMR}

At first glance, weak local realism (wLR) is seen to be a stronger
assumption than weak macroscopic realism (wMR), meaning it is a more
restrictive (less convincing) assumption. However, if the time $t_{f}$
is carefully specified, we show that wLR can be justified by wMR.

Consider the system at time $t_{f}$ after the unitary rotations $U_{\theta}^{A}$
and $U_{\phi}^{B}$ that determine the measurement settings, say $\theta$
and $\phi$, respectively at $A$ and $B$. At this stage, or later,
in the measurement process, there is a coupling of each local system
to a macroscopic meter, via an interaction $H_{M}$. The final state
after coupling is of the form
\begin{eqnarray}
|\psi_{M}\rangle & = & c_{1}|p_{+}\rangle_{A}|\uparrow\rangle_{\theta}|p_{-}\rangle_{B}|\downarrow\rangle_{\phi}+c_{2}|p_{-}\rangle_{A}|\downarrow\rangle_{\theta}|p_{+}\rangle_{B}|\uparrow\rangle_{\phi}\nonumber \\
 &  & +c_{3}|p_{+}\rangle_{A}|\uparrow\rangle_{\theta}|p_{+}\rangle_{B}|\uparrow\rangle_{\phi}+c_{4}|p_{-}\rangle_{A}|\downarrow\rangle_{\theta}|p_{-}\rangle_{B}|\downarrow\rangle_{\phi}\nonumber \\
\label{eq:macro-p}
\end{eqnarray}
where $c_{i}$ are probability amplitudes, and $|p_{+}\rangle_{A/B}$
and $|p_{-}\rangle_{A/B}$ are macroscopic states for the pointer
of the meter, indicating Pauli spin outcomes of $+1$ and $-1$ respectively,
at $A$ and $B$. We see that $|\psi_{M}\rangle$ is a macroscopic
superposition state. Weak macroscopic realism implies predetermined
values $\lambda_{M}^{A}$ and $\lambda_{M}^{B}$ for the outcomes
of measurements on the meter systems $-$ the pointers are already
in some kind of definite state that will indicate the result of the
measurement to be either ``spin up'' or ``spin down''.

In view of the correlation, it can then be argued that the systems
$A$ and $B$ (which may be microscopic) are similarly specified to
be in states with a definite outcome for the final measurement of
spin components $\theta$ and $\phi$, respectively. The interaction
$H_{M}$ is reversible, and hence the definition of wLR can be rephrased
to apply to the system at the time $t_{f}$ where it is assumed the
stage of the measurement that couples each system to a meter has already
occurred, just after or in association with the unitary interactions
$U_{\theta}^{A}$ and $U_{\phi}^{B}$. Due to the reversibility of
$H_{M}$, this does not change the results of the paper.

\section{EPR-Bohm and GHZ paradoxes}

\subsection{Bohm's version of the EPR paradox}

Bohm generalized the EPR paradox to spin measurements by considering
two spatially separated spin $1/2$ particles prepared in the Bell
state \cite{Bohm,bell-1969}
\begin{eqnarray}
|\psi_{B}\rangle & = & \frac{1}{\sqrt{2}}(|\uparrow\rangle_{z}|\downarrow\rangle_{z}-|\downarrow\rangle_{z}|\uparrow\rangle_{z}).\label{eq:bell}
\end{eqnarray}
The particles and their respective sites are denoted by $A$ and $B$.
Here $|\uparrow\rangle_{z}$ and $|\downarrow\rangle_{z}$ are the
eigenstates of the $z$ component $\sigma_{z}$ of the Pauli spin
$\overrightarrow{\sigma}=(\sigma_{x},\sigma_{y},\sigma_{z})$, with
eigenvalues $+1$ and $-1$ respectively. We use the standard notation,
where the first and second states of the product $|\uparrow\rangle|\downarrow\rangle$
refer to the states of particle $A$ and particle $B$ respectively.
The spin operators for the two particles are distinguished by superscripts
e.g. $\sigma_{z}^{A}$ and $\sigma_{z}^{B}$.

\subsubsection{A two-spin version}

From (\ref{eq:bell}), it is clear that the outcomes of spin-$z$
measurements on each particle are anticorrelated. Similarly, we may
measure the component $\sigma_{y}$ of each particle. To predict the
outcomes, we transform the state into the $y$ basis, noting the transformation
\begin{eqnarray}
|\uparrow\rangle_{y} & = & \frac{1}{\sqrt{2}}(|\uparrow\rangle_{z}+i|\downarrow\rangle_{z})\nonumber \\
|\downarrow\rangle_{y} & = & \frac{1}{\sqrt{2}}(|\uparrow\rangle_{z}-i|\downarrow\rangle_{z})\label{eq:transy-1-1}
\end{eqnarray}
where $|\uparrow\rangle_{y}$ and $|\downarrow\rangle_{y}$ are the
eigenstates of $\sigma_{y}$, with respective eigenvalues $+1$ and
$-1$. Hence we find $|\uparrow\rangle_{z}=(|\uparrow\rangle_{y}+|\downarrow\rangle_{y})/\sqrt{2}$
and $|\downarrow\rangle_{z}=-i(|\uparrow\rangle_{y}-|\downarrow\rangle_{y})/\sqrt{2}$.
The state becomes in the new basis
\begin{eqnarray}
|\psi_{B}\rangle & = & \frac{i}{\sqrt{2}}(|\uparrow\rangle_{y}|\downarrow\rangle_{y}-|\downarrow\rangle_{y}|\uparrow\rangle_{y}).\label{eq:bell-state-y}
\end{eqnarray}
Denoting the respective measurements at each site by $\sigma_{y}^{A}$
and $\sigma_{y}^{B}$, we see that the spin-$y$ outcomes at $A$
and $B$ are also anticorrelated.

An EPR-Bohm paradox follows from the following argument (Figure \ref{fig:Sketch-bohm-test}).
By making a measurement of $\sigma_{z}^{B}$ on particle $B$, the
outcome for the measurement $\sigma_{z}^{A}$ on particle $A$ is
known with certainty. EPR present their Assertions of local realism
(LR) \cite{epr}, summarized in Section II.A.1. Invoking EPR Assertion
LR (2) that there is no disturbance to system $A$ due to the measurement
at $B$, EPR's Assertion LR (1) therefore implies system $A$ can
be ascribed a hidden variable $\lambda_{z}^{A}$ \cite{mermin-ghz},
which predetermines the outcome for the measurement $\sigma_{z}^{A}$
should that measurement be performed. However, the outcomes at $A$
and $B$ are also anticorrelated for measurements of $\sigma_{y}$
at both sites. The assumption of EPR's premises therefore ascribes
two hidden variables $\lambda_{z}^{A}$ and $\lambda_{y}^{A}$ to
the system $A$, which simultaneously predetermine the outcome of
either $\sigma_{z}$ or $\sigma_{y}$ at $A$, should either measurement
be performed. This description is not compatible with any quantum
wavefunction $|\psi\rangle$ for the spin $1/2$ system $A$. The
conclusion is that if EPR's local realism is valid, then quantum mechanics
gives an incomplete description of physical reality.

The above conclusions draw on the assumption that the subsystem $A$
is described quantum mechanically as a spin $1/2$ system. For such
a system, the Pauli spin variances defined by $(\Delta\sigma_{i})^{2}=\langle\sigma_{i}^{2}\rangle-\langle\sigma_{i}\rangle^{2}$
satisfy the uncertainty relation $(\Delta\sigma_{x})^{2}+(\Delta\sigma_{y})^{2}+(\Delta\sigma_{z})^{2}\geq2$
\cite{hofmann-take}. Since $(\Delta\sigma_{z})^{2}\leq1$, this implies
\cite{hofmann-take}
\begin{equation}
(\Delta\sigma_{y})^{2}+(\Delta\sigma_{z})^{2}\geq1.\label{eq:uncer-spin}
\end{equation}
For a quantum state description of the system $A$, the values of
$\sigma_{y}$ and $\sigma_{z}$ cannot be simultaneously precisely
defined. A realisation has been given for two spin $1/2$ particles
(photons) which showed near-perfect correlation for both of two orthogonal
spins (orthogonal linear polarizations), for a subensemble where both
photons are detected \cite{aspect-Bohm}.

\subsubsection{Three-spin version}

A stricter argument not dependent on the assumption of a spin $1/2$
system is possible, if the experimentalist can measure the correlation
of all three spin components \cite{Bohm,epr-rmp,bohm-test-uncertainty}.
Consider the spin-$x$ measurements $\sigma_{x}^{A}$ and $\sigma_{x}^{B}$.
The eigenstates of $\sigma_{x}$ are 
\begin{eqnarray}
|\uparrow\rangle_{x} & = & \frac{1}{\sqrt{2}}(|\uparrow\rangle_{z}+|\downarrow\rangle_{z})\nonumber \\
|\downarrow\rangle_{x} & = & \frac{1}{\sqrt{2}}(|\uparrow\rangle_{z}-|\downarrow\rangle_{z}).\label{eq:transx-1-1-2}
\end{eqnarray}
The state (\ref{eq:bell}) becomes in the spin-$x$ basis
\begin{eqnarray}
|\psi_{B}\rangle & = & \frac{1}{\sqrt{2}}(|\downarrow\rangle_{x}|\uparrow\rangle_{x}-|\uparrow\rangle_{x}|\downarrow\rangle_{x}).\label{eq:bell-state-y-1}
\end{eqnarray}
The spin-$x$ outcomes at $A$ and $B$ are also anticorrelated.
According to the EPR premises, it is therefore possible to assign
a hidden variable $\lambda_{x}^{A}$ to the subsystem $A$ that predetermines
the outcome of the measurement $\sigma_{x}^{A}$.  Hence, local realism
implies that the system $A$ would at any time be described by three
precise values, $\lambda_{x}^{A}$, $\lambda_{y}^{A}$ and $\lambda_{z}^{A}$,
which predetermine the outcomes of measurements $\sigma_{x}$, $\sigma_{y}$
and $\sigma_{z}$ respectively. Each of $\lambda_{x}$, $\lambda_{y}$
and $\lambda_{z}$ has the value $+1$ or $-1$ . Since always $|\lambda_{z}^{A}|=1$,
such a hidden variable description cannot be given by a local quantum
state $|\psi\rangle$ of $A$, as this would be a violation of the
quantum uncertainty relation 
\begin{equation}
\Delta\sigma_{x}\Delta\sigma_{y}\geq|\langle\sigma_{z}\rangle|,\label{eq:hup}
\end{equation}
which applies to all quantum states. Hence, we arrive at an EPR paradox,
where local realism implies an inconsistency with the completeness
of quantum mechanics.

\subsection{GHZ paradox}

The Greenberger-Horne-Zeilinger (GHZ) argument shows that local realism
can be falsified, if quantum mechanics is correct \cite{ghz-1,mermin-ghz,ghz-amjp,clifton-ghz}.
The GHZ state
\begin{eqnarray}
|\psi_{GHZ}\rangle & = & \frac{1}{\sqrt{2}}(|\uparrow\rangle_{z}|\uparrow\rangle_{z}|\uparrow\rangle_{z}-|\downarrow\rangle_{z}|\downarrow\rangle_{z}|\downarrow\rangle_{z})\label{eq:ghz-1}
\end{eqnarray}
involves three spatially separated spin $1/2$ particles, $A$, $B$
and $C$. We denote the Pauli spin measurement $\sigma_{\theta}$
at the site $J\in\{A,B,C\}$ by $\sigma_{\theta}^{J}$. Consider measurements
of $\sigma_{x}$ at each site. To obtain the predicted outcomes, we
rewrite in the spin-$x$ basis. The GHZ state becomes
\begin{eqnarray}
|\psi_{GHZ}\rangle & = & \frac{1}{\sqrt{2}}(|\downarrow\rangle_{x}|\uparrow\rangle_{x}|\uparrow\rangle_{x}+|\uparrow\rangle_{x}|\downarrow\rangle_{x}|\uparrow\rangle_{x}\nonumber \\
 &  & +|\uparrow\rangle_{x}|\uparrow\rangle_{x}|\downarrow\rangle_{x}+|\downarrow\rangle_{z}|\downarrow\rangle_{z}|\downarrow\rangle_{z}).\label{eq:ghz-xxx}
\end{eqnarray}
From this we see $\langle\sigma_{x}^{A}\sigma_{x}^{B}\sigma_{x}^{C}\rangle=-1.$
Now we also consider the measurement $\sigma_{x}^{A}\sigma_{y}^{B}\sigma_{y}^{C}$
on the system in the GHZ state. The GHZ state in the spin-$y$
basis is:
\begin{eqnarray}
|\psi_{GHZ}\rangle & = & \frac{1}{\sqrt{2}}(|\uparrow\rangle_{x}|\uparrow\rangle_{y}|\uparrow\rangle_{y}+|\downarrow\rangle_{x}|\downarrow\rangle_{y}|\uparrow\rangle_{y}\nonumber \\
 &  & +|\downarrow\rangle_{x}|\uparrow\rangle_{y}|\downarrow\rangle_{y}+|\uparrow\rangle_{x}|\downarrow\rangle_{y}|\downarrow\rangle_{y}).\label{eq:ghz-xyy}
\end{eqnarray}
This shows $\langle\sigma_{x}^{A}\sigma_{y}^{B}\sigma_{y}^{C}\rangle=1.$
Similarly, $\langle\sigma_{y}^{A}\sigma_{x}^{B}\sigma_{y}^{C}\rangle=1$
and $\langle\sigma_{y}^{A}\sigma_{y}^{B}\sigma_{x}^{C}\rangle=1$.

The GHZ argument is well known. The outcome for $\sigma_{x}$ at $A$
can be predicted with certainty by performing measurements $\sigma_{x}^{B}$
and $\sigma_{x}^{C}$. The measurements do not disturb the system
$A$ because the measurements at $A$ and those at $B$ and $C$ are
spacelike-separated events. Similarly, the outcome for $\sigma_{y}^{A}$
can be predicted, without disturbing the system $A$, by measurements
of $\sigma_{y}^{B}$ and $\sigma_{x}^{C}$. Hence, there exist hidden
variables $\lambda_{x}^{A}$ and $\lambda_{y}^{A}$ that can be simultaneously
ascribed to system $A$, these variables predetermining the outcome
for measurements $\sigma_{x}^{A}$ and $\sigma_{y}^{A}$ at $A$.
The variables assume the values of $+1$ or $-1$. A similar argument
can be made for particles $B$ and $C$. The contradiction with EPR's
local realism arises because the product $\lambda_{x}^{A}\lambda_{x}^{B}\lambda_{x}^{C}$
must equal $-1$, in order that the prediction $\langle\sigma_{x}^{A}\sigma_{x}^{B}\sigma_{x}^{C}\rangle=-1$
holds. Similarly, $\lambda_{x}^{A}\lambda_{y}^{B}\lambda_{y}^{C}=\lambda_{y}^{A}\lambda_{x}^{B}\lambda_{y}^{C}=\lambda_{y}^{A}\lambda_{y}^{B}\lambda_{x}^{C}=1$
in order that the prediction $\langle\sigma_{x}^{A}\sigma_{y}^{B}\sigma_{y}^{C}\rangle=\langle\sigma_{y}^{A}\sigma_{x}^{B}\sigma_{y}^{C}\rangle=\langle\sigma_{y}^{A}\sigma_{y}^{B}\sigma_{x}^{C}\rangle=1$
holds. Yet, we see algebraically that $\lambda_{x}^{A}\lambda_{x}^{B}\lambda_{x}^{C}=\lambda_{x}^{A}\lambda_{x}^{B}\lambda_{x}^{C}(\lambda_{y}^{B})^{2}(\lambda_{y}^{A})^{2}(\lambda_{y}^{C})^{2}$,
and hence
\begin{eqnarray}
\lambda_{x}^{A}\lambda_{x}^{B}\lambda_{x}^{C} & = & (\lambda_{x}^{A}\lambda_{y}^{B}\lambda_{y}^{C})(\lambda_{x}^{B}\lambda_{y}^{A}\lambda_{y}^{C})(\lambda_{x}^{C}\lambda_{y}^{B}\lambda_{y}^{A})\nonumber \\
 & = & 1\label{eq:ghx-contra}
\end{eqnarray}
which gives a complete ``all or nothing'' contradiction. The conclusion
is that local realism does not hold.

\section{An EPR paradox based on weak local realism}

An argument can be formulated that quantum mechanics is incomplete,
based on the wLR premise. The argument follows along similar lines
to the original EPR argument, and is depicted in Figure \ref{fig:Sketch-bohm-test-2}.
The argument applies to the set-up considered by Schrödinger, involving
simultaneous measurements, one direct and the other indirect \cite{sch-epr-exp-atom,s-cat-1935}.

The system at time $t_{0}$ is prepared in the Bell state $|\psi_{B}\rangle$.
Let us choose the measurement setting $\phi\equiv y$ such that the
system $B$ is prepared for the pointer measurement of $\sigma_{y}^{B}$
(denoted $S_{y}^{B}$ on the diagram). From the anticorrelation of
state (\ref{eq:bell-state-y}), the outcome for $\sigma_{y}^{A}$
(i.e. $S_{y}^{A}$) can be predicted with certainty, by measurement
on system $B$. This constitutes Schrödinger's ``indirect measurement'',
of $\sigma_{y}^{A}$ (i.e. $S_{y}^{A}$). Therefore, by Assertions
wLR (1a) and (2a), the system $A$ at time $t_{f_{B}}=t_{f}$, after
$U_{y}^{B}$ has been performed, can be ascribed a definite value
for the variable $\lambda_{y}^{A}$. We note that a further unitary
interaction $U^{A}$ is required at $A$ after time $t_{f}$ so that
the system is prepared for a pointer measurement $\sigma_{y}^{A}$.
Regardless, the final outcome is already determined at time $t_{f_{B}}$
by the value of $\lambda_{y}^{A}$. This inferred variable is depicted
as $\lambda_{y}^{A}$ in black in Figure \ref{fig:Sketch-bohm-test-2}.

However, at the time $t_{f}$,  the system $A$ is \emph{itself }prepared
for a pointer measurement of $\sigma_{z}^{A}$ (i.e. $S_{z}^{A}$).
Hence, by Assertion wLR (1b) and (2b), there is a hidden variable
$\lambda_{z}^{A}$ that predetermines the value for the measurement
$\sigma_{z}^{A}$, should it be performed. This constitutes Schrödinger's
``direct measurement'', of $\sigma_{z}^{A}$ (i.e. $S_{z}^{A}$).
This variable is depicted as $\lambda_{z}^{A}$ in red in Figure \ref{fig:Sketch-bohm-test-2}.
According to the premises, the system $A$ at the time $t_{f}$ therefore
can be ascribed two definite spin values, $\lambda_{z}^{A}$ and $\lambda_{y}^{A}$.
This assignment cannot be given by any localised quantum state for
a spin $1/2$ system, and hence the argument can be put forward similarly
to the original argument that quantum mechanics is incomplete.

In an experiment, it would be demonstrated that the result of $S_{y}^{A}$
can be inferred from the measurement at $B$ with certainty. It would
also be established that system $A$ is given quantum mechanically
as a spin $1/2$ system. A description for a realistic experiment
is given in Appendix D (see also Ref. \cite{sch-epr-exp-atom}).

\paragraph*{Comment:}

The above argument is based on a two-spin version of the EPR-Bohm
argument. The three-spin version could not be formulated using wLR,
because this would require preparation of three pointer measurements,
which is not possible for the bipartite system.  The original three-spin
EPR-Bohm paradox requires the assumption of EPR's LR, which can be
falsified. The two-spin version is based on wLR which has not been
falsified. On the other hand, the two-spin version of the EPR-Bohm
paradox allows a counterargument against the incompleteness of quantum
mechanics: It could be proposed that a local quantum state description
is possible for $A$, but that this description is a complex one,
not describing a spin $1/2$ particle.

\section{Macroscopic EPR-Bohm paradox using cat states}

In this section, we strengthen the EPR argument by presenting cases
where one may invoke macroscopic local realism. We do this by demonstrating
an EPR-Bohm paradox which uses two macroscopically distinct states.
First, we consider where the distinct states are coherent states.
In the second example, the distinct states are a collection of multi-mode
spin states with spins either all ``up', or all ``down''. The unitary
operations $U_{\theta}$ that fix the measurement settings are chosen
to preserve the macroscopic two-state nature of the system, and are
realised by Kerr interactions and CNOT gates.

\subsection{Two-spin EPR-Bohm paradox with coherent states}

We first consider a realisation of the two-spin EPR paradox described
Section III.A.1 using coherent and cat states. This requires macroscopic
spins defined in terms of the two macroscopically distinct coherent
states.

\subsubsection{The initial state, unitary rotations, and definition of macroscopic
spins}

We consider the system to be prepared at time $t_{1}$ in the entangled
cat state \cite{cat-bell-wang-1,cat-det-map}
\begin{equation}
|\psi_{Bell}\rangle=\mathcal{N}(|\alpha\rangle|-\beta\rangle-|-\alpha\rangle|\beta\rangle).\label{eq:bell-epr-1}
\end{equation}
Here $|\alpha\rangle$ and $|\beta\rangle$ are coherent states for
single-mode fields $A$ and $B$, and we take $\alpha$ and $\beta$
to be real, positive and large. $\mathcal{N}=\frac{1}{\sqrt{2}}\{1-\exp(-2\left|\alpha\right|^{2}-2\left|\beta\right|^{2})\}^{-1/2}$
is the normalization constant. The phase of the coherent amplitudes
$\alpha$ and $\beta$ are defined as real relative to an fixed axis,
which is usually defined by a phase specified in the preparation process.
For example, this is may be fixed by the phase of a pump field, as
in the coherent state superpositions generated by nonlinear dispersion
\cite{yurke-stoler-1}.

For each system $A$ and $B$, one may measure the field quadrature
phase amplitudes $\hat{X}_{A}={\color{red}{\color{blue}{\color{black}\frac{1}{\sqrt{2}}}}}(\hat{a}+\hat{a}^{\dagger})$,
$\hat{P}_{A}={\color{red}{\color{blue}{\color{black}\frac{1}{i\sqrt{2}}}}}(\hat{a}-\hat{a}^{\dagger})$,
$\hat{X}_{B}=\frac{1}{\sqrt{2}}(\hat{b}+\hat{b}^{\dagger})$ and $\hat{P}_{A}={\color{red}{\color{blue}{\color{black}\frac{1}{i\sqrt{2}}}}}(\hat{a}-\hat{a}^{\dagger})$,
which are defined in a rotating frame \cite{yurke-stoler-1}. The
boson destruction mode operators for modes $A$ and $B$ are denoted
by $\hat{a}$ and $\hat{b}$. As $\alpha$ $\rightarrow\infty$, the
probability distribution $P(X_{A})$ for the outcome $X_{A}$ of the
measurement $\hat{X}_{A}$ consists of two well-separated Gaussians
which can be associated with the distributions for the coherent states
$|\alpha\rangle$ and $|-\alpha\rangle$. (Any central component due
to interference vanishes for large $\alpha$, $\beta$). Hence, the
outcome $X_{A}$ distinguishes between the states $|\alpha\rangle$
and $|-\alpha\rangle$. Similarly, $\hat{X}_{B}$ distinguishes between
the states $|\beta\rangle$ and $|-\beta\rangle$.

We define the outcome of the ``spin'' measurement $\hat{S}^{A}$
to be $S^{A}=+1$ if $X_{A}\geq0$, and $-1$ otherwise. Similarly,
the outcome of the measurement $\hat{S}^{B}$ is $S^{B}=+1$ if $X_{B}\geq0$,
and $-1$ otherwise. The result is identified as the spin of the system
i.e. the qubit value. For each system, the coherent states become
orthogonal in the limit of large $\alpha$ and $\beta$, in which
case the superposition maps to the two-qubit Bell state $|\psi_{Bell}\rangle=\frac{1}{\sqrt{2}}(|+\rangle_{a}|-\rangle_{b}-|-\rangle_{a}|+\rangle_{b})$,
given by (\ref{eq:bell}). At time $t_{1}$, the outcomes $S^{A}$
and $S^{B}$ are anticorrelated.
\begin{figure}[t]
\begin{centering}
\includegraphics[width=1\columnwidth]{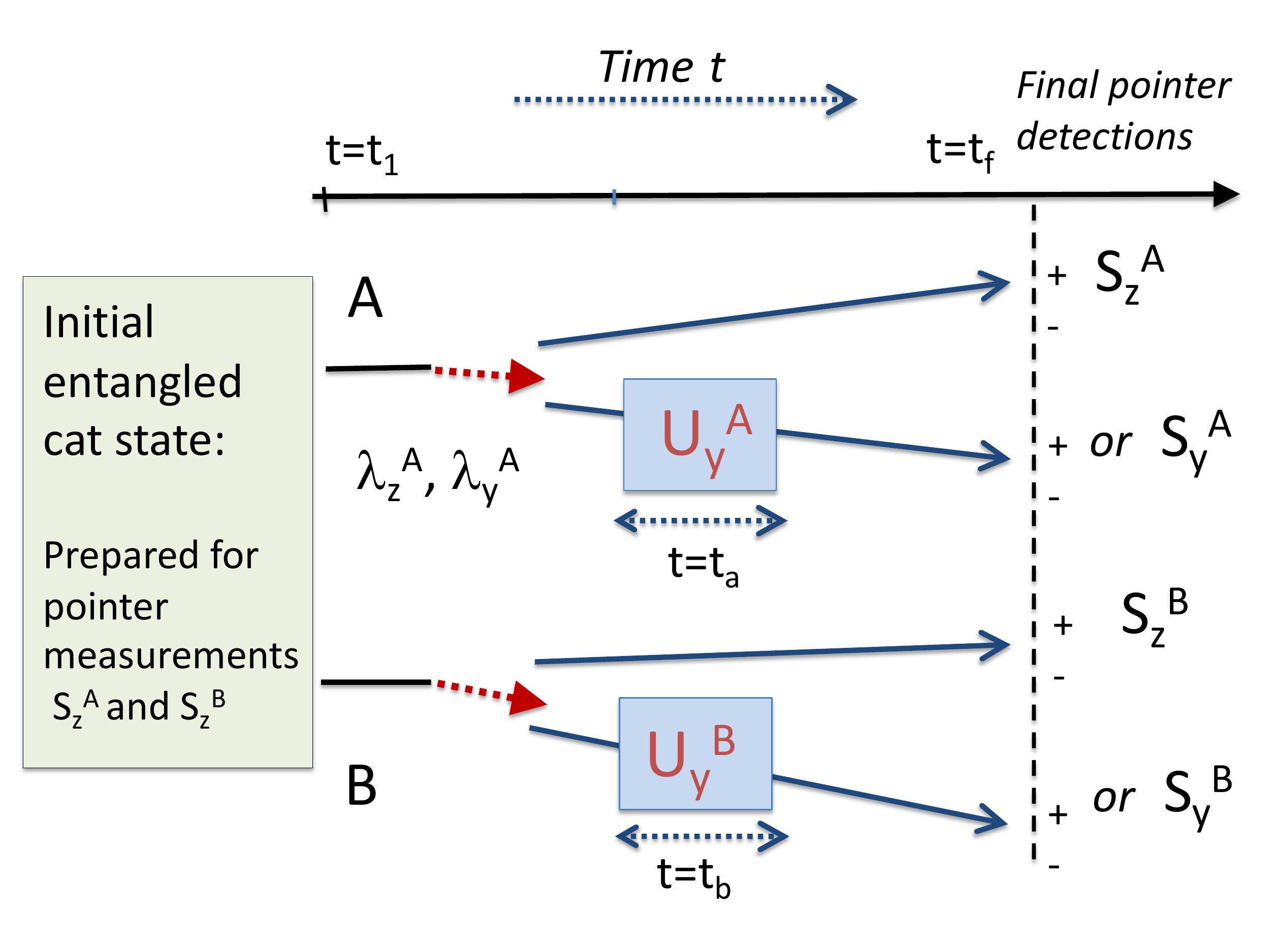}
\par\end{centering}
\caption{Macroscopic version of the EPR-Bohm paradox: The system is prepared
in a cat state, e.g. (\ref{eq:bell-epr-1}), for which the final
outcomes $+$ and $-$ correspond to the macroscopically distinct
amplitudes $\alpha$ and $-\alpha$. At each site $A$ and $B$, a
switch (dashed arrow) allows the independent and random choice to
evolve the systems by $U_{y}$, or not. With no evolution, $S_{z}$
is measured. If the rotation $U_{y}$ takes place, $S_{y}$ is measured.
The outcomes for $S_{z}^{A}$ and $S_{z}^{B}$ (and $S_{y}^{A}$
and $S_{y}^{B}$) are anticorrelated.\label{fig:Sketch-macro-bohm-paradox}}
\end{figure}

In order to realise the EPR-Bohm paradox, it is necessary to identify
the noncommuting spin observables and the appropriate unitary rotations
$U$ at each site required to measure these. For this purpose, we
examine the systems $A$ and $B$ as they evolve independently according
to local transformations $U_{A}(t_{a})$ and $U_{B}(t_{b})$, defined
as
\begin{equation}
U_{A}(t_{a})=e^{-iH_{NL}^{A}t_{a}/\hbar},\ \ U_{B}(t_{b})=e^{-iH_{NL}^{B}t_{b}/\hbar}\label{eq:state5-1}
\end{equation}
where 
\begin{equation}
H_{NL}^{A}=\Omega\hat{n}_{a}^{k},\ \ H_{NL}^{B}=\Omega\hat{n}_{b}^{k}.\label{eq:H}
\end{equation}
Here, $t_{a}$ and $t_{b}$ are the times of evolution at each site,
$\hat{n}_{a}=\hat{a}^{\dagger}\hat{a}$ and $\hat{n}_{b}=\hat{b}^{\dagger}\hat{b}$,
and $\Omega$ is a constant. We consider $k=2$, noting that $k=4$
allows a Bell test \cite{manushan-bell-cat-lg}. The dynamics of
this evolution is well known \cite{yurke-stoler-1,collapse-revival-bec-2,collapse-revival-super-circuit-1,wright-walls-gar-1}.
If the system $A$ is prepared in a coherent state $|\alpha\rangle$,
then  after a time $t_{a}=\pi/2\Omega$ the state of the system
$A$ becomes \cite{manushan-cat-lg,manushan-bell-cat-lg,macro-bell-lg,yurke-stoler-1,cat-states-super-cond,cat-states-mirrahimi}
\begin{eqnarray}
U_{\pi/4}^{A}|\alpha\rangle & = & e^{-i\pi/4}(|\alpha\rangle+i|-\alpha\rangle)/\sqrt{2}.\label{eq:state3-2}
\end{eqnarray}
Here we define $U_{\pi/4}^{A}=U_{A}(\pi/2\Omega)$. A similar transformation
$U_{\pi/4}^{B}$ is defined at $B$ for $t_{b}=\pi/2\Omega$. This
state is a superposition of two macroscopically distinct states, and
is referred to as a cat state after Schrödinger's paradox \cite{s-cat-1935,scat-rmp-frowis}.
Further interaction for the whole period $t_{a}=2\pi/\Omega$ returns
the system to the coherent state $|\alpha\rangle$.

The macroscopic version of the EPR-Bohm paradox is depicted in Figure
\ref{fig:Sketch-macro-bohm-paradox}. We consider the spin-$1/2$
observables $\hat{S}_{z}=|+\rangle\langle+|-|-\rangle\langle-|$,
$\hat{S}_{x}=|+\rangle\langle-|+|-\rangle\langle+|$ and $\hat{S}_{y}=\frac{1}{i}(|+\rangle\langle-|-|-\rangle\langle+|)$,
defined for orthogonal states $|\pm\rangle$ of a two-level system,
which we also denote by $|\uparrow\rangle_{z}$ and $|\downarrow\rangle_{z}$.
Here, we identify the eigenstates $|\pm\rangle$ of $\hat{S}_{z}^{A}$
($\hat{S}_{z}^{B}$) as the coherent states $|\pm\alpha\rangle$ ($|\pm\beta\rangle$)
respectively, with $\alpha$ and $\beta$ real, and in the limit of
large $\alpha$ and $\beta$ where orthogonality is justified. In
this limit, we define
\begin{eqnarray}
\hat{S}_{z}^{A} & = & |\alpha\rangle\langle\alpha|-|-\alpha\rangle\langle-\alpha|\nonumber \\
\hat{S}_{x}^{A} & = & |\alpha\rangle\langle-\alpha|+|-\alpha\rangle\langle\alpha|\nonumber \\
\hat{S}_{y}^{A} & = & \frac{1}{i}(|\alpha\rangle\langle-\alpha|-|-\alpha\rangle\langle\alpha|)\label{eq:spinbohm-1}
\end{eqnarray}
for system $A$. The scaling corresponds to Pauli spins $\overrightarrow{\sigma}=(\sigma_{x},\sigma_{y},\sigma_{z})$.
The spins $\hat{S}_{z}^{B}$, $\hat{S}_{x}^{B}$ and $\hat{S}_{y}^{B}$
for system $B$ are defined in identical fashion on replacing $\alpha$
with $\beta$.  We omit operator ``hats'', where the meaning
is clear.

\subsubsection{Performing the measurement of spins $S_{z}$ and $S_{y}$}

The EPR-Bohm paradox requires measurement of $S_{z}^{A}$ and $S_{z}^{B}$.
The system in the state (\ref{eq:bell-epr-1}) is \emph{prepared for
the pointer stage of the measurements} of $S_{z}$. This is because
for this system, a measurement of (the sign of) $\hat{X}_{A}$ and
$\hat{X}_{B}$ is all that is required to complete the $S_{z}^{A}$
and $S_{z}^{B}$ measurement. The local measurement constitutes an
optical homodyne, in which the fields are combined with a strong field
across a beamsplitter with a relative phase shift $\vartheta$, followed
by direct detection in the arms of the beam splitter \cite{yurke-stoler-1}.
Here, $\vartheta$ is chosen to measure $\hat{X}_{A}$ ($\hat{X}_{B}$),
the axis so that $\alpha$ ($\beta$) as real. The $\vartheta$ is
defined by the preparation process, usually involving a pump field.

The EPR-Bohm argument also requires measurements of $S_{y}^{A}$ and
$S_{y}^{B}$ on the Bell state (\ref{eq:bell-epr-1}) prepared at
time $t_{1}$ (Figure \ref{fig:Sketch-macro-bohm-paradox}). Here,
it is required to adjust the measurement-setting by applying a local
unitary transformation $U_{y}$ at each site. 

The eigenstates of $\hat{S}_{y}$ are often written in the form $|\uparrow\rangle_{y}=(|\uparrow\rangle_{z}+i|\downarrow\rangle_{z})/\sqrt{2}$
and $|\downarrow\rangle_{y}=(|\uparrow\rangle_{z}-i|\downarrow\rangle_{z})/\sqrt{2}$,
but the normalization can vary by a phase factor. We can abbreviate
as $|\pm\rangle_{y}=\frac{1}{\sqrt{2}}(|\pm\rangle+i|\mp\rangle)$,
denoting $|\uparrow\rangle$ as $|+\rangle$, and $|\downarrow\rangle$
as $|-\rangle$, interchangeably. We choose
\begin{eqnarray}
|\uparrow\rangle_{y} & = & \frac{e^{-i\pi/4}}{\sqrt{2}}(|\uparrow\rangle_{z}+i|\downarrow\rangle_{z})\nonumber \\
|\downarrow\rangle_{y} & = & \frac{e^{-i\pi/4}}{\sqrt{2}}(|\downarrow\rangle_{z}+i|\uparrow\rangle_{z})=\frac{e^{i\pi/4}}{\sqrt{2}}(|\uparrow\rangle_{z}-i|\downarrow\rangle_{z}),\nonumber \\
\label{eq:eig-y-1}
\end{eqnarray}
and will denote the eigenstates at different sites by a subscript.
We have temporarily dropped for convenience the superscripts and subscripts
indicating the $A$ and $B$, since the transformations are local
and apply independently to both sites. It is readily verified that
the $\hat{S}_{y}|\uparrow\rangle_{y}=|\uparrow\rangle_{y}$ and $\hat{S}_{y}|\downarrow\rangle_{y}=|\downarrow\rangle_{y}$
i.e. $\hat{S}_{y}^{A}|\pm\rangle_{y,A}=\pm|\pm\rangle_{y,A}$ and
$\hat{S}_{y}^{B}|\pm\rangle_{y,B}=\pm|\pm\rangle_{y,B}$. 

Now we consider how to perform the measurement of $\hat{S}_{y}$.
As explained in Section II, the first stage of measurement involves
a unitary operation $U_{y}$, giving a transformation to the measurement
basis, so that the system is then prepared for the second stage of
measurement, which is the ``pointer {[}stage of{]} measurement''
of $\hat{S}_{y}$. The pointer stage constitutes a measurement of
the sign $\hat{S}$ of $\hat{X}$, which for large $\alpha$ ($\beta$)
will (after $U_{y}$ has been applied) directly yield the outcome
$\pm1$ for the system prepared in $|\pm\rangle_{y}$.  To establish
$U_{y}$, following the procedure of Eqs. (\ref{eq:bell}-\ref{eq:bell-state-y})
and (\ref{eq:ghz-1}-\ref{eq:ghz-xyy}), any state 
\begin{equation}
|\psi\rangle=c_{+}|\uparrow\rangle_{z}+c_{-}|\downarrow\rangle\label{eq:c}
\end{equation}
written in the $z$ basis can be transformed into the $y$ basis,
by substituting
\begin{eqnarray}
|\uparrow\rangle_{z} & \rightarrow & (e^{i\pi/4}|\uparrow\rangle_{y}+e^{-i\pi/4}|\downarrow\rangle_{y})/\sqrt{2}\nonumber \\
|\downarrow\rangle_{z} & \rightarrow & -i(e^{i\pi/4}|\uparrow\rangle_{y}-e^{-i\pi/4}|\downarrow\rangle_{y})/\sqrt{2},\label{eq:ct}
\end{eqnarray}
This gives 
\begin{equation}
|\psi\rangle=d_{+}|+\rangle_{y}+d_{-}|-\rangle_{y}\label{eq:tr}
\end{equation}
where $d_{\pm}=(c_{\pm}\mp ic_{-})e^{\pm i\pi/4}$. To obtain the
transformed state (\ref{eq:tr}), ready for the pointer stage of measurement
of $S_{y}$, the system \emph{is thus evolved} according to 
\begin{equation}
U_{y}|\psi_{Bell}\rangle\label{eq:s}
\end{equation}
where $U_{y}\equiv U_{\pi/4}^{-1}=U^{-1}(\pi/2\Omega)$. We explain
this result further in the Appendix A for the purpose of clarity.

The $U_{y}$ is the inverse of the transformation $e^{-iH_{NL}t/\hbar}$
where $t=\pi/2\Omega$, given by (\ref{eq:state3-2}). The $U_{y}$
is achieved by evolving the local system for a time $t=-\pi/2\Omega\equiv3\pi/2\Omega$,
since the solutions are periodic.

\paragraph{Comment}

The states $|+\rangle_{y}$ and $|-\rangle_{y}$ refer to the macroscopically
distinct \emph{coherent states} $|\alpha\rangle$ and $|-\alpha\rangle$
defined at the time $t$ \emph{after} the local unitary rotation $U_{y}$
has taken place. This is important in identifying the macroscopic
nature of the paradox. We then see that the premises of weak macroscopic
realism defined in Section II.A.2 will apply (Figure \ref{fig:Sketch-bohm-test-2}).

\subsubsection{EPR-Bohm argument}

The EPR-Bohm argument is as follows (Figure \ref{fig:Sketch-macro-bohm-paradox}).
Consider the system prepared in the Bell state (\ref{eq:bell-epr-1})
at time $t_{1}$ ($\alpha$, $\beta\rightarrow\infty$). One first
measures $S_{z}^{A}$ and $S_{z}^{B}$ for this state at time $t_{1}$.
The anticorrelation of the Bell state means that the result for
$S_{z}^{A}$ at $A$ can be predicted with certainty by the measurement
at $B$.

The EPR-Bohm argument continues, by considering measurements of $S_{y}^{A}$
and $S_{y}^{B}$ on the Bell state (\ref{eq:bell-epr-1}) prepared
at time $t_{1}$ (Figure \ref{fig:Sketch-macro-bohm-paradox}). The
measurement of $S_{y}^{A}$ ($S_{y}^{B}$) is thus made by applying
the local unitary rotation $U_{y}^{A}$ ($U_{y}^{B}$) to the Bell
state prepared at $t_{1}$, followed by a measurement of the sign
of $X_{A}$ ($X_{B}$). The state of the system prepared after the
unitary rotations $U_{y}^{A}$ and $U_{y}^{B}$ is also given as the
Bell state (\ref{eq:bell-epr-1}), which we write as
\begin{equation}
|\psi_{Bell}\rangle_{y,y}=\frac{1}{\sqrt{2}}(|-\rangle_{y}|+\rangle_{y}-|+\rangle_{y}|-\rangle_{y})\label{eq:basis}
\end{equation}
where we have taken $\alpha$, $\beta$ large. As with the original
paradox given in Section III.A.1, as seen by Eq. (\ref{eq:bell-state-y}),
the final measurements of $X_{A}$ and $X_{B}$ therefore reveal an
anticorrelation between $S_{y}^{A}$ and $S_{y}^{B}$, and the result
for $S_{y}^{A}$ can be revealed, with certainty, by measurement of
$S_{y}^{B}$ at site $B$. Here, we note the Comment in the above
section, that $|\pm\rangle_{y}$ are the macroscopically distinct
coherent states $|\alpha\rangle$ and $|-\alpha\rangle$ (or $|\beta\rangle$
and $|-\beta\rangle$) that are realised at the time $t_{f}$ in Figure
\ref{fig:Sketch-macro-bohm-paradox}, which corresponds to the time
after the transformation $U_{y}$ has been carried out at each location.

The Bohm-EPR argument continues. The correlation between the spins
enables an experimentalist at $B$ to determine with certainty either
$S_{z}^{A}$ or $S_{y}^{A}$, for the system prepared at $t_{1}$,
by choosing the suitable measurement at the site $B$. Assuming EPR's
local realism, this implies that both the spins of system $A$ are
predetermined with certainty. Following along the lines of the two-spin
paradox of Section III.A.1, this constitutes Bohm's EPR paradox, because
it is not possible to define a local quantum state $\varphi$ for
the spin $1/2$ system $A$ at the time $t_{1}$ with simultaneously
specified values for both $\hat{S}_{z}^{A}$ and $\hat{S}_{y}^{A}$.
The paradox is the inconsistency between macroscopic local realism
and the completeness of quantum mechanics.

In the gedanken experiment, it is assumed that the system $A$ is
described quantum mechanics as a spin $1/2$ system, which is valid
as $\alpha\rightarrow\infty$, where the two coherent states $|\alpha\rangle$
and $|-\alpha\rangle$ are orthogonal. In the same limit, the spin
outcomes for $\hat{S}_{z}^{A}$ and $\hat{S}_{z}^{B}$, and also for
$\hat{S}_{y}^{A}$ and $\hat{S}_{y}^{B}$, are perfectly anticorrelated,
so that this realisation of the EPR-Bohm paradox strictly follows
in the macroscopic limit, where $\alpha\rightarrow\infty$. Proposals
for finite $\alpha$ that also account for imperfect anticorrelation
of the spins are presented in Appendices C and D.

\subsection{Two- and three-spin paradox with spins and CNOT gates}

A useful realization uses multimode spin states and CNOT gates. This
allows a realisation of both types of EPR-Bohm paradox presented in
Section III.A, the two- and three-spin versions, at an increasingly
macroscopic level depending on the number of modes. Here, because
the spin qubits correspond to macroscopically distinct states, the
paradoxes will reveal an inconsistency between macroscopic local realism
and the completeness of quantum mechanics.

By analogy with the microscopic example of Section III.A.2, the three-spin
paradox requires a transformation $U_{x}$ at each site, where (apart
from phase factors)
\begin{eqnarray}
U_{x}^{-1}|\uparrow\rangle & \rightarrow & \frac{1}{\sqrt{2}}(|\uparrow\rangle+|\downarrow\rangle)\nonumber \\
U_{x}^{-1}|\downarrow\rangle & \rightarrow & \frac{1}{\sqrt{2}}(|\uparrow\rangle-|\downarrow\rangle),\label{eq:trans-cat-x-1}
\end{eqnarray}
as well as that for $U_{y}$, given as
\begin{eqnarray}
U_{y}^{-1}|\uparrow\rangle_{z} & \rightarrow & \frac{1}{\sqrt{2}}(|\uparrow\rangle+i|\downarrow\rangle)\nonumber \\
U_{y}^{-1}|\downarrow\rangle_{z} & \rightarrow & \frac{1}{\sqrt{2}}(|\uparrow\rangle-i|\downarrow\rangle).\label{eq:trans-cat-y-2-2}
\end{eqnarray}
The important step is to find a Hamiltonian that gives a realisation
of $U_{x}$ and $U_{y}$. In the previous section, for the cat-states
involving the coherent states, a transformation $U_{y}$ was specified
but not for $U_{x}$. We note that cat-state superpositions $|\alpha\rangle\pm|-\alpha\rangle$
can be created using conditional measurements \cite{cat-state-phil,cat-states-super-cond,cat-det-map},
and open dissipative systems \cite{cat-states-wc,cat-even-odd-transient,transient-cat-states-leo,cat-dynamics-ry,cats-hach,cat-states-mirrahimi,cat-det-map,cat-states-super-cond}.
However, we prefer to use simple unitary transformations. A
realisation based on NOON states is given in Appendix B.

A realisation can be achieved using an array of spins. The qubits
of (\ref{eq:trans-cat-y-2-2}) become the macroscopically distinct
states $|\uparrow\rangle\equiv|\uparrow\rangle^{\otimes N}$ and $|\downarrow\rangle\equiv|\downarrow\rangle^{\otimes N}$,
for large $N$, so that the initial Bell state (\ref{eq:bell}) becomes
the two-site GHZ state
\begin{equation}
|\psi_{Bell}\rangle_{z,z}=\frac{1}{\sqrt{2}}(|\uparrow\rangle_{z,A}^{\otimes N}|\uparrow\rangle_{z,B}^{\otimes N}-|\downarrow\rangle_{z,A}^{\otimes N}|\downarrow\rangle_{,B}^{\otimes N}).\label{eq:three-site-ghz-1}
\end{equation}
The premises of macroscopic realism can be applied to the macroscopically
distinct states. Here, $|\uparrow\rangle_{z,J}^{\otimes N}=\prod_{k=1}^{N}|\uparrow\rangle_{J,k}$
where $|\uparrow\rangle_{J,k}$ is the eigenstate of the Pauli spin
$\sigma_{z}^{k}$ for the mode labelled $k$ at site $J$, the collection
of modes $k=1,..N$ forming the system labelled $J$. The $|\uparrow\rangle_{J}^{\otimes N}$
and $|\downarrow\rangle_{J}^{\otimes N}$ represent macroscopically
distinct states, with collective Pauli spin values of $N$ or $-N$,
and are eigenstates of the spin product $S_{z}^{J}=\prod_{k=1}^{N}\sigma_{z}^{k}$.

In order to realise the paradox, the transformations $U$ needed at
each site $J$ are, for $U_{x}$ and $U_{y}$, of the form (\ref{eq:trans-cat-x-1})-(\ref{eq:trans-cat-y-2-2}),
but where we replace $|\uparrow\rangle\equiv|\uparrow\rangle^{\otimes N}$
and $|\downarrow\rangle\equiv|\downarrow\rangle^{\otimes N}$.  Generally,
one can first consider how to achieve
\begin{equation}
|\uparrow\rangle^{\otimes N}\rightarrow\cos\frac{\theta}{2}|\uparrow\rangle^{\otimes N}+e^{i\vartheta}\sin\frac{\theta}{2}|\downarrow\rangle^{\otimes N}.\label{eq:transtheta}
\end{equation}
Following the experiment described in \cite{IBM-macrorealism-1},
the unitary transformations $U_{x}$ and $U_{y}$ are made in two
steps.

The first step is a rotation on the single-mode spin $|\uparrow\rangle_{1}\equiv\begin{pmatrix}1\\
0
\end{pmatrix}$, $|\downarrow\rangle_{1}\equiv\begin{pmatrix}0\\
1
\end{pmatrix}$, given by the unitary matrix $U_{\theta,\vartheta}=\begin{pmatrix}\cos\frac{\theta}{2} & -\sin\frac{\theta}{2}\\
e^{i\vartheta}\sin\frac{\theta}{2} & e^{i\vartheta}\cos\frac{\theta}{2}
\end{pmatrix}$ where $\vartheta=0$ or $\pi/2$, which transforms the spin as
\begin{eqnarray}
|\uparrow\rangle_{1} & \rightarrow & U_{\theta,\vartheta}|\uparrow\rangle_{1}=\cos\frac{\theta}{2}|\uparrow\rangle_{1}+e^{i\vartheta}\sin\frac{\theta}{2}|\downarrow\rangle_{1}\nonumber \\
|\downarrow\rangle_{1} & \rightarrow & U_{\theta,\vartheta}|\downarrow\rangle_{1}=-\sin\frac{\theta}{2}|\uparrow\rangle_{1}+e^{i\vartheta}\cos\frac{\theta}{2}|\downarrow\rangle_{1}.\label{eq:uni}
\end{eqnarray}
Here, we drop the subscript $J$ representing the site, for notational
simplicity. Choosing $\theta=\pi/2$ gives the starting point for
the transformation $U_{x}$ or $U_{y}$ at each site, with $\vartheta=0$
or $\pi/2$ respectively.

A common physical realisation of the spin qubit involves two polarisation
modes: $|\uparrow\rangle\equiv|1,0\rangle$ and $|\downarrow\rangle\equiv|0,1\rangle$
defined for two modes $a_{\pm}$ as in Appendix B. The transformation
$U_{\theta,\vartheta}$ can then be achieved with a polarizing beam
splitter, with mode transformations ($\hat{a}_{\pm}$ are boson operators
defining the modes)
\begin{eqnarray}
\hat{c}_{+} & = & \hat{a}_{+}\cos\theta-\hat{a}_{-}\sin\theta\nonumber \\
e^{i\vartheta}\hat{c}_{-} & = & \hat{a}_{+}\sin\theta+\hat{a}_{-}\cos\theta.\label{eq:bs}
\end{eqnarray}
The $\hat{c}_{\pm}$ are boson operators for the outgoing modes emerging
from the beam splitter. The interaction is described by the Hamiltonian
$H=i\hbar k(\hat{a}_{+}\hat{a}_{-}^{\dagger}-\hat{a}_{+}^{\dagger}\hat{a}_{-})$
where $\theta=kt$, for $\vartheta=0$. The addition of a $\vartheta=\pi/2$
phase shift (or not) relative to the two outputs gives the mode transformations
with the dependence on $\vartheta=0$ or $\pi/2$. If the input is
$|\uparrow\rangle$, the output state is
\begin{eqnarray}
|1,0\rangle_{in} & = & \hat{a}_{+}^{\dagger}|0\rangle\nonumber \\
 & = & \cos\theta|1,0\rangle_{out}+e^{i\vartheta}\sin\theta|0,1\rangle_{out}.\label{eq:st}
\end{eqnarray}
 If the input is $|\downarrow\rangle$, the output is found according
to
\begin{eqnarray}
|0,1\rangle_{in} & = & \hat{a}_{-}^{\dagger}|0\rangle\nonumber \\
 & = & -\sin\theta|1,0\rangle_{out}+e^{i\vartheta}\cos\theta|0,1\rangle_{out}\label{eq:bsstates}
\end{eqnarray}
which gives a starting point for the transformation $U_{x}$ (where
$\vartheta=0$) and $U_{y}$ (where $\vartheta=\pi/2$) at each site
$J$.

The second step of the transformations $U_{x}$ and $U_{y}$ involves
a sequence of CNOT gates. Consider the example of two qubits, with
the initial state $|00\rangle\equiv|\uparrow\rangle|\uparrow\rangle$.
The transformation $U_{\theta,\vartheta}$ on the first qubit evolves
the state into:
\begin{align}
U_{\theta,\vartheta}|\uparrow\rangle|\uparrow\rangle & =\cos\frac{\theta}{2}|\uparrow\rangle|\uparrow\rangle+e^{i\vartheta}\sin\frac{\theta}{2}|\downarrow\rangle|\uparrow\rangle.\label{eq:rot}
\end{align}
The subsequent CNOT gate then flips the second (target) qubit to $|1\rangle\equiv|\downarrow\rangle$
if the first (control) qubit is $|1\rangle$. For $n>2$, the CNOT
gates will be performed between the first qubit and all other qubits.
This gives 
\begin{align}
U_{\theta,\vartheta}|\uparrow\rangle^{\otimes N} & =\cos\frac{\theta}{2}|\uparrow\rangle^{\otimes N}+e^{i\vartheta}\sin\frac{\theta}{2}|\downarrow\rangle^{\otimes N}.\label{eq:rot-1}
\end{align}
In this way, the transformations (\ref{eq:trans-cat-x-1})-(\ref{eq:trans-cat-y-2-2})
for $U_{x}$ and $U_{y}$ can be achieved macroscopically (for large
$N$) for each site.

In the two-spin experiment, either $U_{y}$ or $U_{z}$ is selected
at each site, in order to measure $S_{y}^{J}$ or $S_{z}^{J}$. We
specify that the initial state $|\psi_{Bell}\rangle_{z,z}$ (Eq. (\ref{eq:three-site-ghz-1}))
has been prepared for the pointer measurement of $S_{z}^{J}$. This
means that a direct detection of the qubit value (such as a direct
detection of a photon in the mode $a_{+}$ or $a_{-}$) is all that
is required to complete the measurement of $S_{z}^{J}$.

The experiment of \cite{IBM-macrorealism-1} used the IBM quantum
computer to perform the operations with $N=2-6$, enabling a test
of macrorealism.  In a macroscopic realisation, similar operations
have been performed using Rydberg atoms, for $N\sim20$ \cite{omran-cats}.

The analysis given in Section V.A above follows for this example,
on replacing the macroscopically distinct states $|\alpha\rangle$
and $|-\alpha\rangle$ with $|\uparrow\rangle^{\otimes N}$ and $|\downarrow\rangle^{\otimes N}$.
One can define the macroscopic spins and the eigenstates $|\pm\rangle_{y}$
and $|\pm\rangle_{x}$ of $S_{y}$ and $S_{x}$ similarly. Following
the Comment in Section V.A, the states after the transformations $U_{y}$
and $U_{x}$ are thus superpositions of the two macroscopically distinct
states $|\uparrow\rangle^{\otimes N}$ and $|\downarrow\rangle^{\otimes N}$
(for large $N$), which are prepared for the pointer measurement.
Hence, the premises of weak macroscopic realism defined in Section
II.A.2 apply (Figure \ref{fig:Sketch-bohm-test-2}). The application
of the premises macroscopic local realism and deterministic macroscopic
realism is explained in the next section. The macroscopic paradoxes
map onto the microscopic ones discussed in Section III, and the predictions
for the correlations follow accordingly.

\section{Conclusions from the macroscopic EPR-Bohm paradox}

The EPR-Bohm paradoxes of Section V involving cat states give a stronger
version of the EPR argument. The predetermined values for the spins
are macroscopically distinct, being the amplitudes $\alpha$ and $-\alpha$,
or else the collective Pauli spin values of $N$ and $-N$. Two types
of EPR paradox based on macroscopic realism can be put forward. The
first is based on MLR (or deterministic macroscopic realism), which
can be falsified. The second is based on weak macroscopic realism.

\subsection{EPR paradox based on deterministic macroscopic realism}

Macroscopic local realism (MLR) is EPR's local realism when applied
to the system of Figure \ref{fig:Sketch-macro-bohm-paradox}, where
the outcomes $+$ and $-$ imply \emph{macroscopically distinct} states
for the system defined at time $t_{1}$ (Section II.A.2). The Locality
Assertion LR (2) becomes more convincing, since any disturbance of
$A$ due to the measurement at $B$ would then require a macroscopic
change of the state at $A$. The application of EPR's local realism
to the system for both measurements $S_{z}$ and $S_{y}$ leads to
the conclusion that system $A$ is described simultaneously by both
hidden variables $\lambda_{z}^{A}$ and $\lambda_{y}^{A}$ at the
time $t_{1}$ (and for the three-spin paradox, similarly for $S_{x}$).
The macroscopic paradox therefore indicates inconsistency between
MLR and the completeness of quantum mechanics. It is important however,
that we justify the application of MLR to \emph{both} measurements.

According to the definition given by Leggett and Garg, the premises
of MLR and dMR require identification of \emph{two macroscopically
distinct states that the system at that time ``has available to it''}.
At the time $t_{1}$, the systems of Section V.A and V.B are superpositions
of two states $|\uparrow\rangle_{z}$ and $|\downarrow\rangle_{z}$
($|\alpha\rangle$ and $|-\alpha\rangle$, or $|\uparrow\rangle^{\otimes N}$
or $|\downarrow\rangle^{\otimes N}$) that can be regarded as macroscopically
distinct (for large $\alpha$ and $N$). These states are prepared
for the pointer measurement $S_{z}$.

The application of the premises to the set-up also requires that the
states $|\uparrow\rangle_{y}$ and $|\downarrow\rangle_{y}$ distinguished
by the measurement $\hat{S}_{y}$ be regarded as macroscopically distinct
at this time $t_{1}$. The eigenstates can be represented as superpositions
e.g. $|\uparrow\rangle_{z}\pm|\downarrow\rangle_{z}$ of the macroscopically
distinct states say, $|\alpha\rangle$ and $|-\alpha\rangle$, at
this time. It is argued that the superpositions represented by the
different probability amplitudes are macroscopically distinct, because
two basis states are. The distinction can be made macroscopic in terms
of the pointer basis, by applying a unitary transformation $U_{y}$
which does not involve amplification.

The macroscopic versions of the EPR-Bohm paradox can be based on deterministic
macroscopic realism alone, defined in Section II.A.2. The term ``deterministic''
is used, because in the context of the EPR-Bohm setup, the premise
implies that the system (at the time $t_{1}$) is simultaneously specified
by both hidden variables, $\lambda_{z}^{A}$ and $\lambda_{y}^{A}$.
 The outcome for measurement of either $S_{z}^{A}$ or $S_{y}^{A}$
at $A$ is considered predetermined, without regard to the measurement
apparatus, as in classical mechanics.

We now argue that the assumption of deterministic macroscopic realism
(dMR) is equivalent to that of macroscopic local realism (MLR) for
the macroscopic EPR set-up. In the EPR set-up, for any macroscopic
spin $S_{\theta}^{A}$ ($\theta\equiv z,y$), one may determine which
of two macroscopically distinct states the macroscopic system $A$
is in, without disturbing system $A$, by performing a spacelike separated
measurement on $B$. Thus, for the Bohm example where we realize anticorrelated
outcomes between $A$ and $B$, dMR is implied by MLR. The converse
is also true. The premise of dMR is that system $A$ already be in
a state with predetermined value $\lambda$ for the spin $S^{A}$
(whether $S_{z}^{A}$ or $S_{y}^{A}$), prior to the measurement being
performed. The locality assumption at a macroscopic level is naturally
part of the definition of dMR: The value of $\lambda^{A}$ cannot
be affected by measurements performed on a spacelike separated system
$B$. The anticorrelation allows determination of the predetermined
value for $A$, given a measurement at $B$. Thus, it follows that
dMR implies MLR. Hence, we use the terms dMR and MLR interchangeably
in this paper.

We mention that Leggett and Garg motivated tests of macroscopic realism
\cite{legggarg-1}. However, in order to establish a test, the additional
assumption of noninvasive measurability was introduced for single
systems. Therefore, reports of violations of Leggett-Garg inequalities
(e.g. \cite{asadian-lg,emary-review,IBM-macrorealism-1,leggett-garg-uola,NSTmunro-1-1,manushan-cat-lg})
do not imply falsification of macroscopic realism, but rather of the
combined premises of macrorealism.

The EPR-Bohm paradox for cat states thus illustrates inconsistency
between dMR (or MLR) and the notion that quantum mechanics is a complete
theory.  However, dMR (and MLR) can be falsified by violations of
Bell inequalities for cat states \cite{manushan-bell-cat-lg,manushan-cat-lg,macro-bell-lg,macro-bell-jeong}.
We show in Section VII that dMR (and MLR) can also be falsified in
a macroscopic GHZ set-up.

\subsection{An EPR paradox based on weak macroscopic realism}

\begin{figure}[t]
\begin{centering}
\includegraphics[width=1\columnwidth]{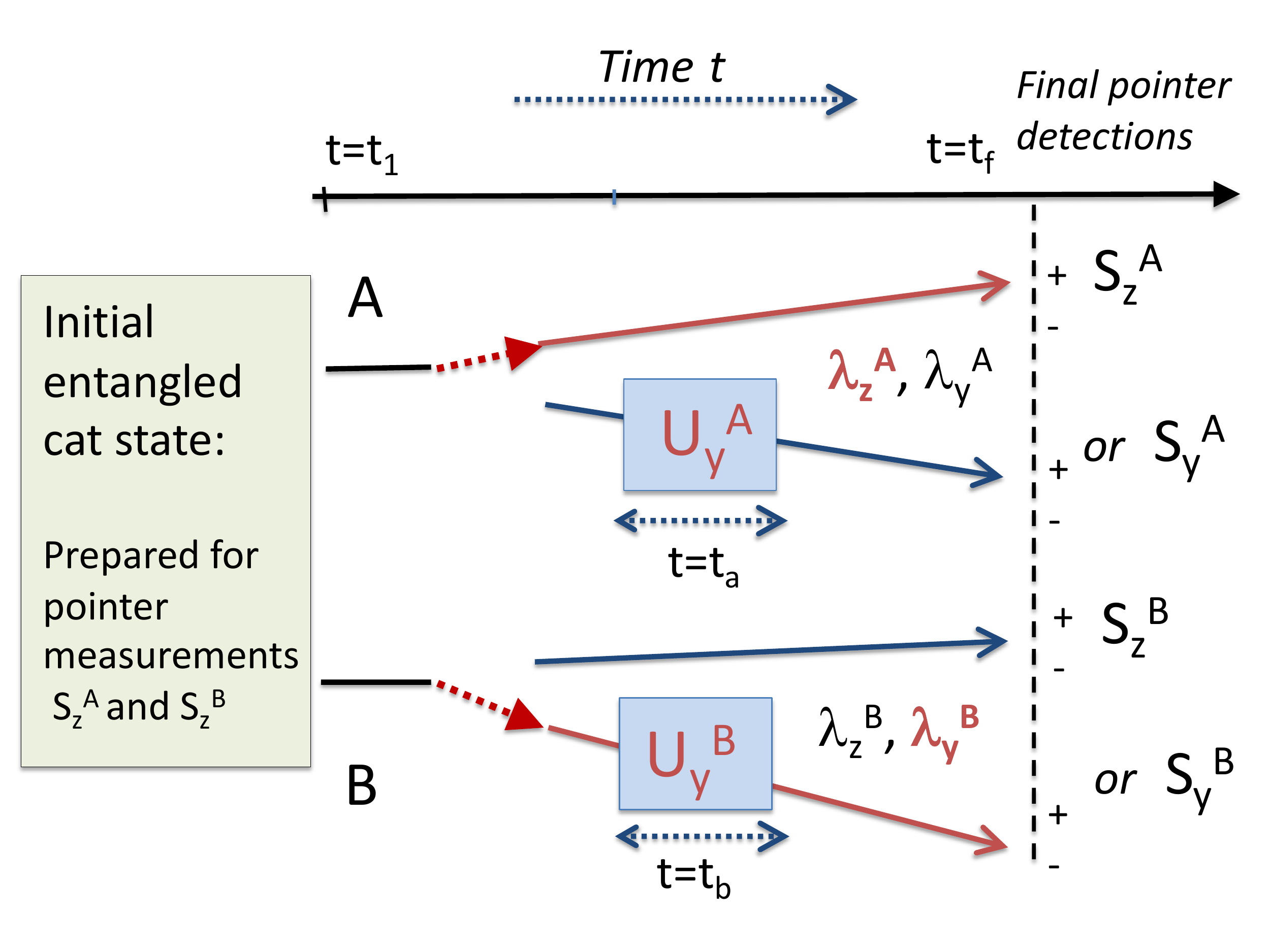}
\par\end{centering}
\caption{A macroscopic EPR-Bohm paradox based on weak macroscopic realism:
The system is as described for Figure \ref{fig:Sketch-macro-bohm-paradox}.
At the time $t_{f}$, the sketch depicts the choice to measure $S_{z}^{A}$
and $S_{y}^{B}$. Hence, wMR ascribes to system $A$ at time $t_{f}$
a predetermined value $\lambda_{z}^{A}$ for the outcome for $S_{z}^{A}$,
since the system has been prepared for a pointer measurement of $S_{z}^{A}$.
The final outcomes for $S_{z}^{A}$ and $S_{z}^{B}$ (and $S_{y}^{A}$
and $S_{y}^{B}$) are anticorrelated. Hence, wMR also ascribes a hidden
variable $\lambda_{y}^{A}$ for system $A$ that predetermines the
outcome $S_{y}^{A}$. \label{fig:Sketch-macro-bohm-paradox-1}}
\end{figure}

It is also possible to make an argument for the incompleteness of
quantum mechanics, based on the premise of weak macroscopic realism
(wMR) (Figure \ref{fig:Sketch-macro-bohm-paradox-1}). The macroscopic
paradox follows along the same lines as that for weak local realism,
given in Section IV, except that the outcomes $+$ and $-$ for the
spins $S_{z}$ and $S_{y}$ can be shown to correspond to macroscopically
distinct states for the system measured at time $t_{f}$. The EPR-Bohm
paradox based on wMR is stronger than that based on deterministic
macroscopic realism, or macroscopic local realism, because the assumption
of wMR is weaker and is \emph{not} falsified by the GHZ or Bell predictions.

The argument for the inconsistency between wMR and the notion of the
completeness of quantum mechanics is illustrated in Figure \ref{fig:Sketch-macro-bohm-paradox-1}.
At time $t_{f}$ the system $B$ has undergone the evolution $U_{y}^{B}$
to prepare the system $B$ for the pointer measurement of $S_{y}^{B}$,
whereas system $A$ is prepared for the pointer measurement of $S_{z}^{A}$.
The premise wMR asserts by Assertion (1b) that at time $t_{f}$, a
value $\lambda_{z}^{A}$ predetermines the outcome for $S_{z}^{A}$
at $A$, and similarly, a value $\lambda_{y}^{B}$ predetermines the
outcome for $S_{y}^{B}$ at $B$. By Assertion (1a), because of the
anticorrelation between $A$ and $B$, the value of $\lambda_{y}^{A}=-\lambda_{y}^{B}$
also predetermines the outcome for $S_{y}^{A}$ at $A$, at the time
$t_{f}$ (even though a further unitary rotation at $A$ would be
necessary to carry out the measurement).  Thus, wMR asserts that
the system $A$ at the time $t_{f}$ can be simultaneously assigned
values $\lambda_{z}^{A}$ and $\lambda_{y}^{A}$ predetermining the
results of measurements $S_{z}^{A}$ and $S_{y}^{A}$. Hence, there
is an EPR paradox.

The quantum correlations of the macroscopic EPR-Bohm, Bell and GHZ
paradoxes are consistent with wMR, because the systems are prepared
for a pointer measurement of $S_{z}$ at one time $t_{1}$, and then
can be prepared for a pointer measurement of $S_{y}$ at a later time
$t_{f}$, after further unitary rotations $U_{A}$ or $U_{B}$. The
hidden variables for the EPR-Bohm paradox are tracked in Figure \ref{fig:Sketch-macro-bohm-paradox-1}.
 We note wMR does not assert that at the time $t_{f}$, the value
of an arbitrary third measurement $S_{\theta}$ is predetermined prior
to the unitary rotation, since that rotation $U$ has not been performed
at site $A$ or $B$. 

For a bipartite system, it is the introduction of a third measurement
setting that leads to the falsification of dMR, as evident by the
Bell tests which require three or more different measurement settings
\cite{bell-1969,manushan-bell-cat-lg}. A falsification is possible
for dMR, because the premise dMR asserts that the system $A$ (or
$B$) has simultaneously predetermined values for the outcomes of
all pointer measurements, at the time $t_{1}$, prior to unitary dynamics
$U$ that finalizes the choice of measurement setting.

The ideal experiment realizing the paradox based on wMR would establish
that the outcome for $S_{y}^{A}$ can be inferred with certainty from
the measurement at $B$. It would also establish that systems $A$
and $B$ are in a superposition (or mixture) of the two relevant macroscopically
distinct states. It is also necessary to demonstrate that the system
$A$ is a spin $1/2$ system, as in Eqs. (\ref{eq:spinbohm-1}), e.g.
demonstrating both measurements $S_{z}^{A}$ and $S_{y}^{A}$ and
the relation (\ref{eq:uncer-spin}). Conditions for a realistic experiment
are given in Appendix D.

\section{GHZ cat gedanken experiment}

The GHZ argument outlined in Section III.B becomes macroscopic when
the spins $|\uparrow\rangle$ and $|\downarrow\rangle$ correspond
to macroscopically distinct states. The macroscopic set-up begins
with the preparation at time $t_{1}$ of the GHZ state
\begin{eqnarray}
|\psi_{GHZ}\rangle & = & \frac{1}{\sqrt{2}}(|\uparrow\rangle_{z,A}|\uparrow\rangle_{z,B}|\uparrow\rangle_{z,C}-|\downarrow\rangle_{z,A}|\downarrow\rangle_{z,B}|\downarrow\rangle_{z,C})\nonumber \\
\label{eq:ghz-cat}
\end{eqnarray}
where $|\uparrow\rangle_{z,J}\equiv|\uparrow\rangle_{z,J}^{\otimes N}$
and $|\downarrow\rangle_{z,J}\equiv|\downarrow\rangle_{z,J}^{\otimes N}$,
defined in Section V.B, are eigenstates of $S_{z}^{J}=\prod_{k=1}^{N}\sigma_{z}^{k}$
with eigenvalues $1$ and $-1$ respectively. Here, $J\equiv A,B,C$
denotes the site. As explained in Section V.B, the system is prepared
at $t_{1}$ for a pointer measurement of $S_{z}^{A}S_{z}^{B}S_{z}^{C}$.
One then considers the measurements of $S_{x}^{A}S_{x}^{B}S_{x}^{C}$
and $S_{x}^{A}S_{y}^{B}S_{y}^{C}$. By analogy with the microscopic
example, this involves applying the transformations $U_{x}$ or $U_{y}$
given by (\ref{eq:trans-cat-x-1})-(\ref{eq:trans-cat-y-2-2}) at
each site. After the interactions $U_{x}^{A}$, $U_{x}^{B}$ and $U_{x}^{C}$,
the system is prepared for the pointer measurement of $S_{x}^{A}S_{x}^{B}S_{x}^{C}$.
The state in the new basis is
\begin{eqnarray}
|\psi_{GHZ}\rangle & = & \frac{1}{2}(|\downarrow\rangle_{x,A}|\uparrow\rangle_{x,B}|\uparrow\rangle_{x,C}+|\uparrow\rangle_{x,A}|\downarrow\rangle_{x,B}|\uparrow\rangle_{x,C}\nonumber \\
 &  & +|\uparrow\rangle_{x,A}|\uparrow\rangle_{x,B}|\downarrow\rangle_{x,C}+|\downarrow\rangle_{x.A}|\downarrow\rangle_{x.B}|\downarrow\rangle_{x,C})\nonumber \\
\label{eq:cat-in-spinx-1-1}
\end{eqnarray}
The product of the spins is $S_{x}^{A}S_{x}^{B}S_{x}^{C}=-1$. 

If we evolve the state (\ref{eq:ghz-cat}) with $U_{x}^{A}$, $U_{y}^{B}$
and $U_{y}^{C}$, the system is prepared for a pointer measurement
of $S_{x}^{A}S_{y}^{B}S_{y}^{C}$. In the new basis, 
\begin{eqnarray}
|\psi_{GHZ}\rangle & = & \frac{1}{4}(|\uparrow\rangle_{y,A}|\uparrow\rangle_{y,B}|\uparrow\rangle_{y,C}+|\downarrow\rangle_{y,A}|\downarrow\rangle_{y,B}|\uparrow\rangle_{y,C}\nonumber \\
 &  & +|\downarrow\rangle_{y,A}|\uparrow\rangle_{y,B}|\downarrow\rangle_{y,C}+|\uparrow\rangle_{y,A}|\downarrow\rangle_{y,B}|\downarrow\rangle_{y,C}).\nonumber \\
\label{eq:cat-spin-y}
\end{eqnarray}
Always, $S_{x}^{A}S_{y}^{B}S_{y}^{C}=1$. Similarly, we consider $S_{y}^{A}S_{y}^{B}S_{x}^{C}$
and $S_{y}^{A}S_{x}^{B}S_{y}^{C}$, and arrive at the GHZ contradiction,
as for the microscopic case. The unitary interactions $U_{x}$ and
$U_{y}$ were shown possible using CNOT gates in Section V.B.
\begin{figure}[t]
\begin{centering}
\includegraphics[width=1\columnwidth]{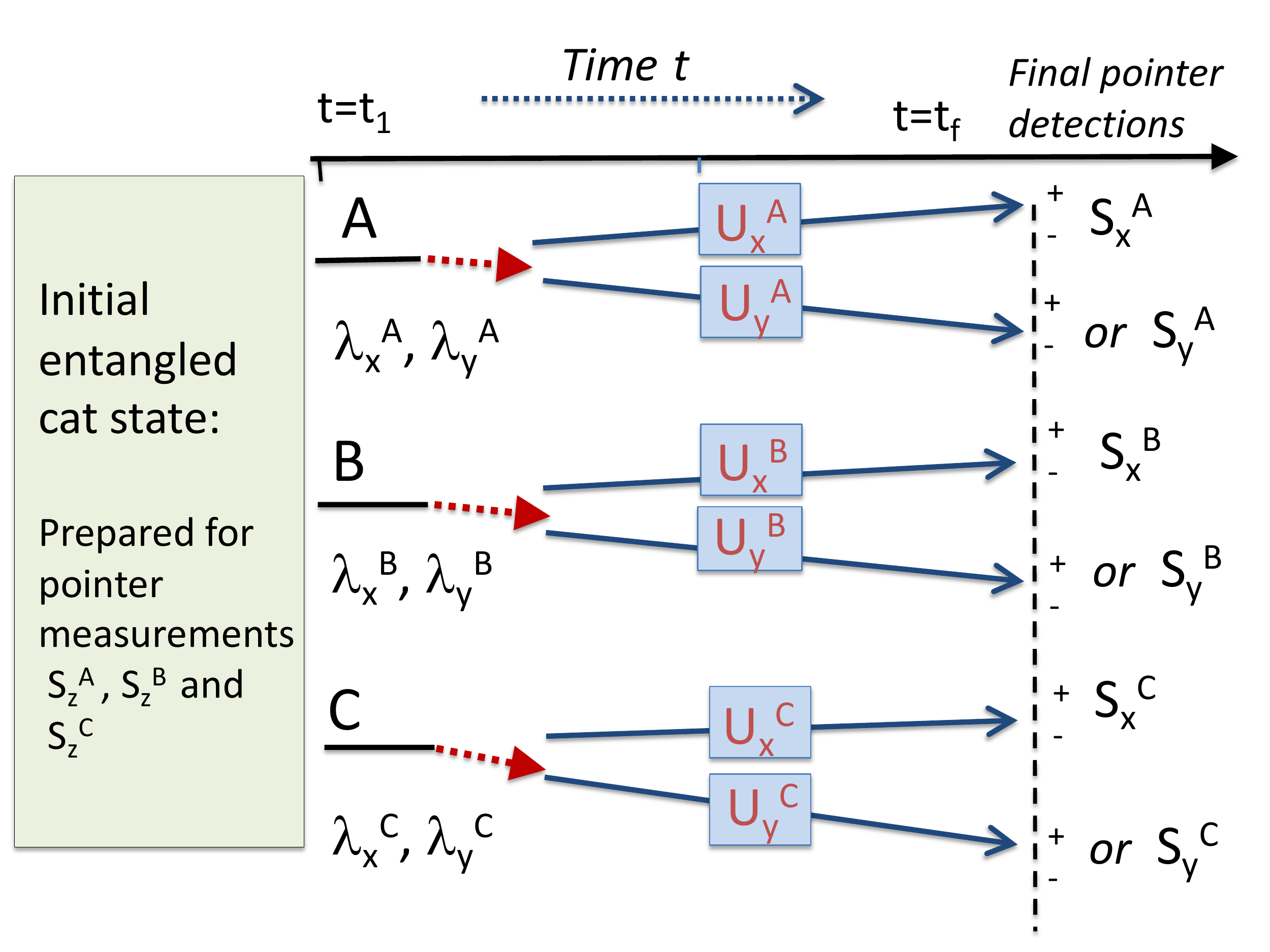}
\par\end{centering}
\caption{Set-up for the GHZ paradox with cat states. The outcome of $S_{x}$
(and $S_{y}$) at each of the sites $A$, $B$ and $C$ can be predicted
with certainty, by choosing certain spacelike separated measurements
at the other two sites. EPR's (macroscopic) local realism implies
the outcomes are predetermined by variables $\lambda_{x}$ and $\lambda_{y}$
at the time $t_{1}$, as indicated on the diagram, which gives the
GHZ contradiction. The hidden variables can also be deduced from the
premise of deterministic macroscopic realism. The GHZ contradiction
is a falsification of deterministic macroscopic realism. \label{fig:Sketch-ghz-cat}}
\end{figure}

\section{Conclusions from the GHZ-cat gedanken experiment}

The macroscopic GHZ set-up enables a falsification of the MLR premises,
and hence is a stronger version of the GHZ experiment. This is because
the states $|\uparrow\rangle_{z,J}\equiv|\uparrow\rangle_{z,J}^{\otimes N}$
and $|\downarrow\rangle_{z,J}\equiv|\downarrow\rangle_{z,J}^{\otimes N}$
are macroscopically distinct for large $N$, and the transformations
$U_{x}$ and $U_{y}$ given by (\ref{eq:trans-cat-x-1})-(\ref{eq:trans-cat-y-2-2})
create superpositions of the macroscopically distinct states. Applying
the justification given in Section VI.A that the eigenstates $|\uparrow\rangle_{y,J}$
and $|\downarrow\rangle_{y,J}$ (and $|\uparrow\rangle_{x,J}$ and
$|\downarrow\rangle_{x,J}$) are also macroscopically distinct, the
hidden variables $\lambda_{x}^{A}$, $\lambda_{y}^{A}$, $\lambda_{x}^{B}$,
$\lambda_{y}^{B}$, $\lambda_{x}^{C}$, $\lambda_{y}^{C}$ defined
for the GHZ system in Section III.B are deduced based on MLR. MLR
asserts that the spacelike measurement at $B$ or $C$ cannot induce
a macroscopic change to the system $A$. This is a weaker assumption
than local realism, which rules out all changes. The GHZ contradiction
explained in Section VI.A falsifies MLR.

\subsection{Falsification of deterministic macroscopic realism}

The GHZ paradox as applied to the cat states is also a falsification
of deterministic macroscopic realism (dMR). We may present the GHZ
paradox directly from the premise of dMR. The premise dMR asserts
that the system $A$ (as it exists at time $t_{1}$) can be ascribed
a hidden variable $\lambda_{\theta}$, the value of which gives the
outcome of the macroscopic spin $S_{\theta}^{A}$, should that measurement
be performed, because the eigenstates of $S_{\theta}^{A}$ are assumed
macroscopically distinct. The premise dMR asserts that the value
of $\lambda_{\theta}$ is not affected by measurements on spacelike
separated systems. One may determine which of the two macroscopically
distinct states (given by $\lambda_{\theta}=1$ or $-1$) the system
$A$ is in, by the measurements on $B$ and $C$. Deterministic macroscopic
realism asserts that the hidden variable $\lambda_{\theta}$ applies
to the system $A$, prior to the selection of the measurement settings
at $B$ and $C$. The set-up is as in Figure \ref{fig:Sketch-ghz-cat},
where a switch controls whether $S_{x}^{A}$ or $S_{y}^{A}$ will
be inferred at $A$, by measuring either $S_{x}^{B}S_{x}^{C}$, or
$S_{x}^{B}S_{y}^{C}$. The argument is that the measurement set-up
at $B$ and $C$ does not disturb the outcome at $A$, and hence both
values, $\lambda_{x}^{A}$ and $\lambda_{y}^{A}$, are simultaneously
determined at $A$, at the time $t_{1}$.

The GHZ paradox thus demonstrates that dMR will fail, assuming the
paradox can be experimentally realised in agreement with quantum predictions.
This is a strong result, giving an ``all or nothing'' contradiction
with dMR. Other macroscopic versions of the GHZ paradox \cite{ghz-macro-1,ghz-macro-2,ghz-macro-3,mermin-inequality}
refer to multidimensional systems, and usually do not address the
macroscopic distinction between the spin states. The falsification
of MLR and dMR undermines the macroscopic EPR-Bohm argument for the
incompleteness of quantum mechanics, given in Section V, which is
based on the assumption that these premises are valid.

\subsection{Consistency of GHZ quantum predictions with wMR and wLR}

The conclusion that macroscopic realism does not hold would be a startling
one. This motivates consideration of the less restrictive definition
of weak macroscopic realism (wMR), defined in Section II.B. The premise
wMR can be applied to the GHZ set-up to show there is \emph{no inconsistency
of wMR with the quantum predictions}. This is in agreement with previous
work \cite{manushan-bell-cat-lg,delayed-choice-cats}, where consistency
with wMR was shown for  Bell violations using cat states.

\begin{figure}[t]
\begin{centering}
\includegraphics[width=1\columnwidth]{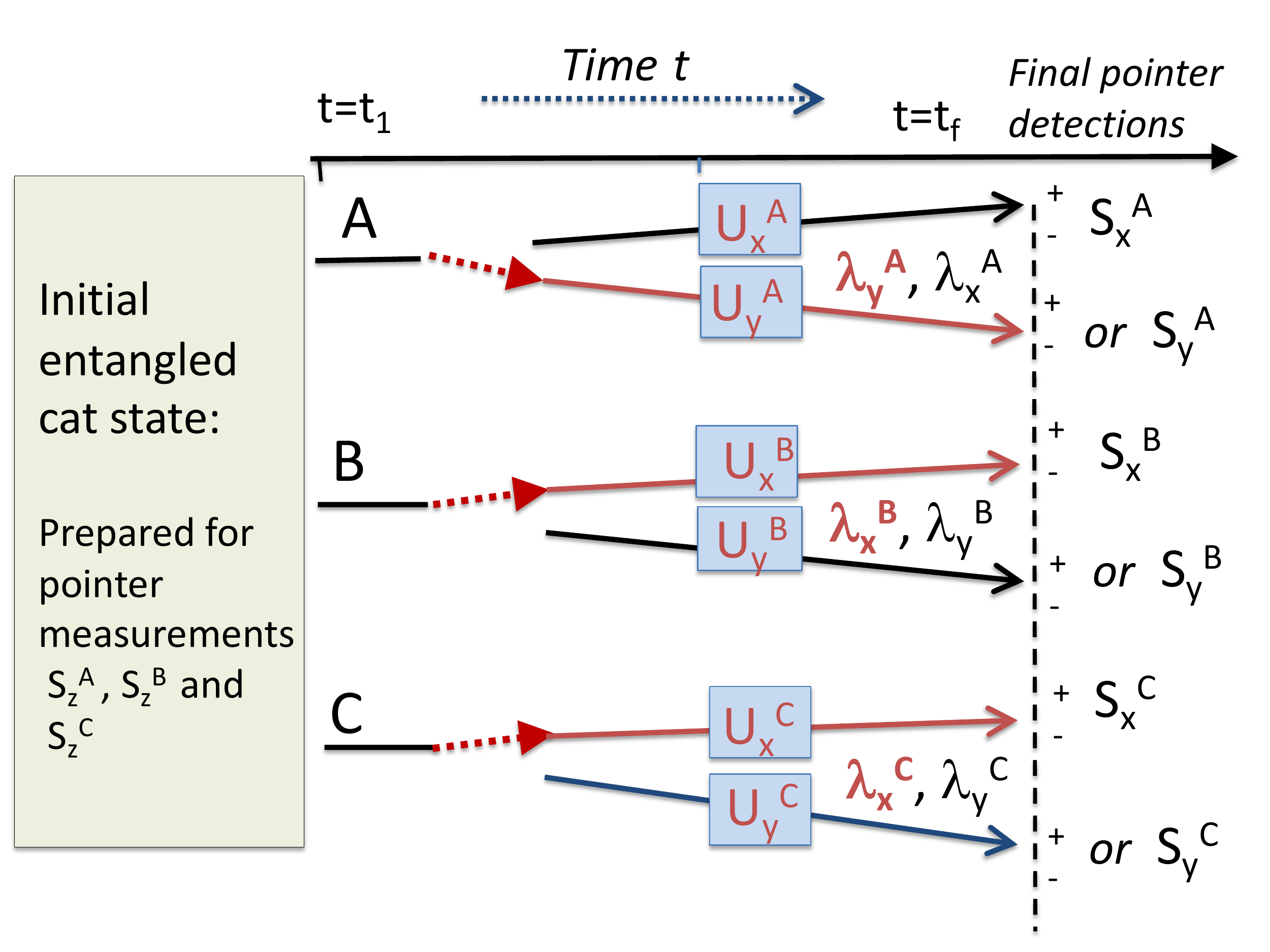}
\par\end{centering}
\caption{Weak macroscopic realism (wMR) applied to the GHZ experiment. The
premise wMR asserts validity of hidden variables for systems at time
$t_{f}$ prepared for the \emph{pointer} measurements. Sketched is
the set-up where $S_{y}$ is measured at $A$, and $S_{x}$ at $B$
and $S_{x}$ at $C$. The wMR premise asserts hidden variables for
the system $A$ at the time $t_{f}$ that predetermine the final outcome
of $S_{y}$ at $A$, and also predetermine the outcome of $S_{x}$
at $A$. This is because the prediction for $S_{x}$ at $A$ can be
given with certainty by the pointer measurements at $B$ and $C$.
Similar logic implies hidden variables that predetermine the outcomes
for both $S_{x}$ and $S_{y}$ for sites $B$ and $C$. However, there
is no contradiction with wMR. This is because the hidden variables
$\lambda_{x}^{A}$, $\lambda_{y}^{B}$ and $\lambda_{y}^{C}$ give
the outcomes of pointer measurements to be made after a further local
unitary interaction $U$, assuming there are no further unitary interactions
at the other sites. \label{fig:Sketch-ghz-cat-weak}}
\end{figure}

To demonstrate the consistency with wMR, we consider state (\ref{eq:ghz-cat})
at time $t_{1}$, and then suppose the systems $B$ and $C$ are prepared
so that pointer measurements of $S_{x}^{B}$ and $S_{x}^{C}$, at
the time $t_{f}$ will yield the outcomes of $S_{x}^{B}$ and $S_{x}^{C}$
(as in Figure \ref{fig:Sketch-ghz-cat-weak}). Weak macroscopic realism
asserts that the systems are each assigned a predetermined value $\lambda_{x}^{B}$
and $\lambda_{x}^{C}$ respectively for the outcomes of those pointer
measurements, at the time $t_{f}$. The premise of wMR also assigns
an inferred value
\begin{equation}
\lambda_{x}^{A}\equiv\lambda_{x,inf}^{A}=\lambda_{x}^{B}\lambda_{x}^{C}\label{eq:lamb1}
\end{equation}
to the system $A$, since the values $\lambda_{x}^{B}$ and $\lambda_{x}^{C}$
enable a prediction with certainty for the outcome of the measurement
$S_{x}^{A}$, if performed at $A$.

It is also the case however that wMR applies directly to $A$. If
the system $A$ undergoes rotation $U_{y}$, as depicted in Figure
\ref{fig:Sketch-ghz-cat-weak}, then it is prepared in a pointer superposition
with respect to $S_{y}^{A}$. Hence the system $A$ is ascribed a
hidden variable $\lambda_{y}^{A}$ to predetermine the outcome $S_{y}^{A}$
based on the pointer preparation of the system A itself.

At first glance, this seems to suggest a GHZ contradiction for wMR.
Suppose one prepares the systems $B$ and $C$ for the pointer measurements
of $S_{x}^{B}$ and $S_{x}^{C}$, at time $t_{f}$ (Figure \ref{fig:Sketch-ghz-cat-weak}).
Hence, for the systems $B$ and $C$ at time $t_{f}$, the outcomes
for $S_{x}^{A}$, $S_{x}^{B}$ and $S_{x}^{C}$ are all predetermined,
and given by variables $\lambda_{x}^{A}\equiv\lambda_{x,inf}^{A}$,
$\lambda_{x}^{B}$ and $\lambda_{x}^{C}$ respectively. Additionally,
one can prepare system $A$ in pointer measurement for $y$, and the
outcome for $S_{y}^{A}$ is also determined (Figure \ref{fig:Sketch-ghz-cat-weak}).
Then, one can infer the values for the outcomes of measurements $S_{y}^{B}$
and $S_{y}^{C}$, should they be performed by carrying out the appropriate
unitary interaction at $B$ and $C$. We have for the inferred values:
\begin{eqnarray}
\lambda_{x,inf}^{A} & = & -\lambda_{x}^{B}\lambda_{x}^{C}\nonumber \\
\lambda_{y,inf}^{C} & = & \lambda_{x}^{B}\lambda_{y}^{A}\nonumber \\
\lambda_{y,inf}^{B} & = & \lambda_{x}^{C}\lambda_{y}^{A}.\label{eq:parameters-reln}
\end{eqnarray}
For each system, the value of either $S_{x}$ or $S_{y}$ is determined
(by the pointer preparation), and the value of the other measurement
is determined, by inference of the other (pointer) values. Hence,
it appears that there is the GHZ contradiction, because it is as though
the outcomes of both $S_{x}$ and $S_{y}$ are determined at each
site (at the same time), and these outcomes are either $+1$ or $-1$,
hence creating the contradiction of Eq. (\ref{eq:ghx-contra}).

However, there is no contradiction with wMR. The value for either
$S_{x}$ or $S_{y}$ (the one that is inferred at each site) will
require a local unitary rotation $U$ (a change of measurement setting)
before the final read-out given by a pointer measurement. The unitary
interaction $U$ occurs over a time interval. The unitary rotation
means that the value $\lambda$ that predetermines the outcome of
the pointer measurement at time $t_{f}$ no longer (necessarily) applies
at the later time, $t_{m}$, after the interaction $U$. The system
at $t_{m}$ is prepared with respect to a different pointer measurement.
Hence, at time $t_{m}$, the earlier predictions of the inferred values
$\lambda_{inf}$ for the other sites no longer apply. The paradox
as arising from Eq. (\ref{eq:ghx-contra}) assumes all values of $\lambda$
apply, to the state at time $t_{f}$.
\begin{figure}[t]
\begin{centering}
\includegraphics[width=1\columnwidth]{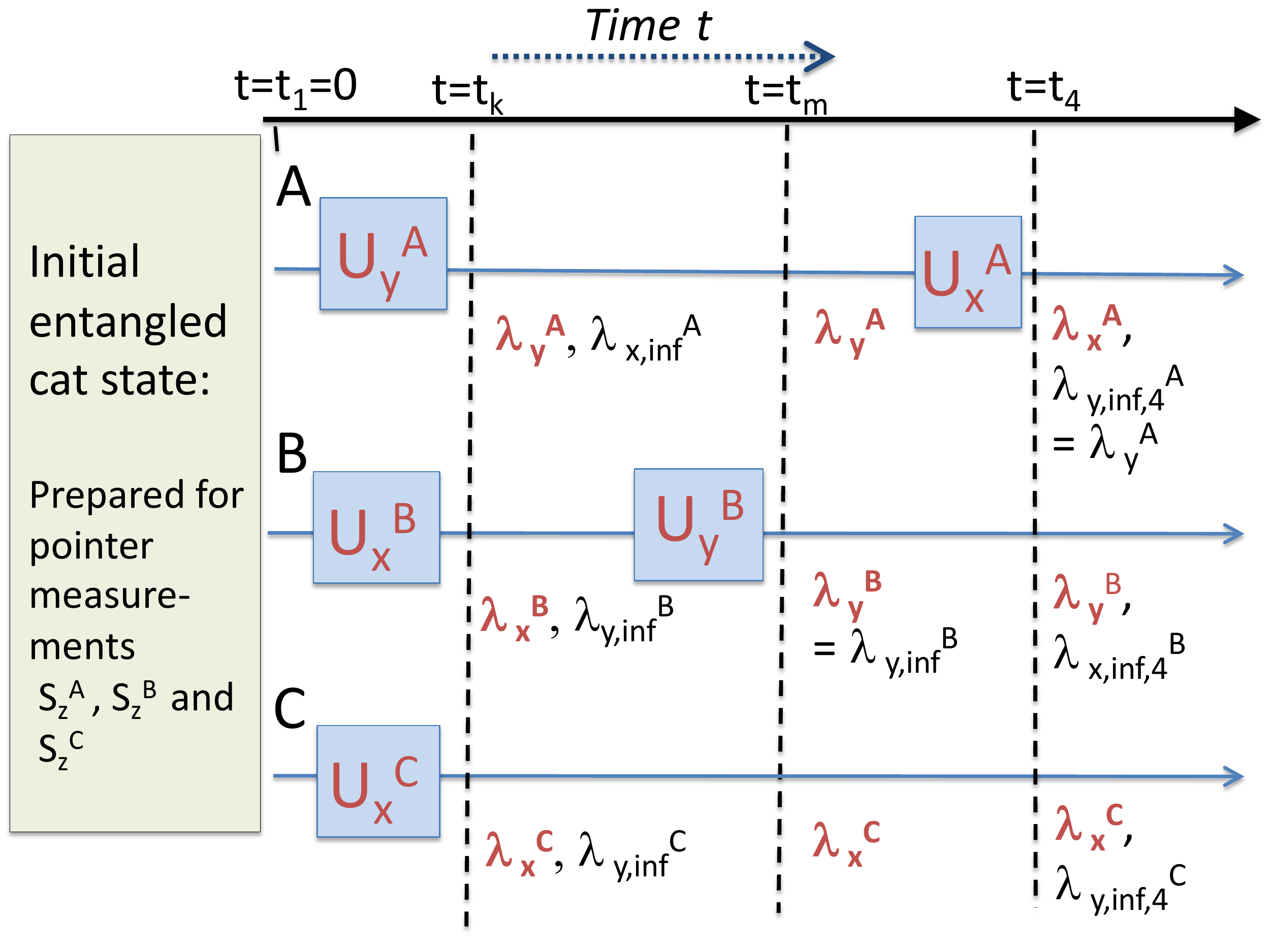}
\par\end{centering}
\caption{Tracking the hidden variables as given by the premise of weak macroscopic
realism (wMR). The dynamics is entirely consistent with the predictions
of wMR, despite there being a GHZ paradox. These variables $\lambda$
give the outcomes for the appropriate measurement if performed. The
variables in bold red are implied by Assertion (1b) of wMR, that the
system is prepared at that time $t_{i}$ at that site, with respect
to the pointer basis. The variables indicated by the subscript $inf$
are deduced by Assertion (1a), that the outcome for the measurement
can be predicted by the pointer measurements at the other sites, at
that time $t_{i}$. The hidden variables that are implied by wMR at
the times $t_{i}$ are indicated beside the dashed vertical line labelled
$t_{i}$.\label{fig:Sketch-ghz-cat-weak-consistency}}
\end{figure}

In Figure \ref{fig:Sketch-ghz-cat-weak-consistency}, we give more
details of the way in which the hidden variables implied by wMR can
be tracked and found consistent with the predictions of quantum mechanics.
Suppose the system is prepared ready for pointer measurements $S_{y}^{A}$,
$S_{x}^{B}$ and $S_{x}^{C}$ at the time $t_{k}$, and the hidden
variables $\lambda_{y}^{A}$, $\lambda_{x}^{B}$ and $\lambda_{x}^{C}$
(in bold red) determine those pointer outcomes. The decision is then
made to measure instead $S_{y}^{A}$, $S_{y}^{B}$ and $S_{x}^{C}$.
This requires a further unitary rotation $\mathbf{U}_{y}^{B}=U_{y}U_{x}^{-1}$
at site $B$. At time $t_{m}$, after $\mathbf{U}_{y}^{B}$ has taken
place, the system is described by a different set of hidden variables,
$\lambda_{y}^{A}$, $\lambda_{y}^{B}$ and $\lambda_{x}^{C}$. The
outcome of the measurement of $S_{y}$ is however determined with
certainty by the pointer measurements for $A$ and $C$, as $\lambda_{y}^{B}=\lambda_{y,inf}^{B}=\lambda_{y}^{A}\lambda_{x}^{C}$.
At time $t_{m}$, we then see that system $B$ is no longer prepared
in a pointer state for $S_{x}$. Hence, at time $t_{m}$, the earlier
value of the inferred result $\lambda_{x,inf}^{A}$ at $A$ (which
depended on $\lambda_{x}^{B}$) is not relevant. A further unitary
rotation $\mathbf{U}_{x}^{A}=U_{x}U_{y}^{-1}$ at $A$ that prepares
the system $A$ for a final pointer measurement $S_{x}$ will not
(necessarily) give the results that applied at time $t_{k}$ (which
was prior to the $U_{y}$ at $B$ taking place). Consider the hidden
variables that are defined (based on the premise of wMR) at this time,
$t_{4}$, after the evolution $\mathbf{U}_{x}^{A}$. At the time $t_{4}$,
the system is ascribed the variables $\lambda_{x}^{A}$, $\lambda_{y}^{B}$
and $\lambda_{x}^{C}$, with $\lambda_{y}^{A}$ also determined, for
a future single unitary transformation at $A$. The outcome for $S_{y}$
at $A$ is predetermined (by the pointer outcomes at $B$ and $C$)
according to wMR, given by
\begin{eqnarray}
\lambda_{y,inf,4}^{A} & = & \lambda_{y}^{B}\lambda_{x}^{C}=\lambda_{y,inf}^{B}\lambda_{x}^{C}\nonumber \\
 & = & \lambda_{x}^{C}\lambda_{y}^{A}\lambda_{x}^{C}\nonumber \\
 & = & \lambda_{y}^{A},\label{eq:lam3}
\end{eqnarray}
which gives consistency with the earlier value at $t_{k}$. However,
the outcome for $S_{x}$ at $B$ is inferred from the pointer values
at time $t_{4}$:
\begin{eqnarray}
\lambda_{x,inf,4}^{B} & = & -\lambda_{x}^{A}\lambda_{x}^{C}.\label{eq:lam4}
\end{eqnarray}
For consistency with the values defined at $t_{k}$, we could propose
$\lambda_{x}^{A}=\lambda_{x,inf}^{A}=-\lambda_{x}^{B}\lambda_{x}^{C}$,
in which case we would obtain $\lambda_{x,inf,4}^{B}=-\lambda_{x}^{A}\lambda_{x}^{C}=\lambda_{x}^{B}(\lambda_{x}^{C})^{2}=\lambda_{x}^{B}$,
giving an apparent consistency with the earlier value. However, the
value of $S_{y}$ at $C$ at time $t_{4}$ is inferred to be
\begin{eqnarray}
\lambda_{y,inf,4}^{C} & = & \lambda_{y}^{B}\lambda_{x}^{A}=\lambda_{y,inf}^{B}\lambda_{x}^{A}\nonumber \\
 & = & (\lambda_{x}^{C}\lambda_{y}^{A})\lambda_{x}^{A}.\label{eq:lam5}
\end{eqnarray}
Now if we propose $\lambda_{x}^{A}=\lambda_{x,inf}^{A}=-\lambda_{x}^{B}\lambda_{x}^{C}$,
we obtain $\lambda_{y,inf,4}^{C}=(\lambda_{x}^{C}\lambda_{y}^{A})(-\lambda_{x}^{B}\lambda_{x}^{C})=-\lambda_{y}^{A}\lambda_{x}^{B}$.
We see here that this is different to the earlier value, $\lambda_{y,inf}^{C}=\lambda_{y}^{A}\lambda_{x}^{B}$.
Hence, it is not possible to gain consistency between wMR and the
values $\lambda$ asserted by the premise of dMR. While dMR is falsified
by the GHZ paradox, we see that the GHZ contradiction does not apply
to wMR.

We note that according to wMR, the value $\lambda_{y}^{A}$ for system
$A$ prepared for the pointer measurement $S_{y}^{A}$, for example,
is not changed by unitary rotations that may take place at $B$ or
$C$ (Figure \ref{fig:Sketch-ghz-cat-weak-consistency}, at time $t_{m}$).
However, if there is a further unitary rotation at $A$, and also
at $B$ (i.e. two unitary rotations, at different sites) the value
$\lambda_{x,inf}^{A}$ for $S_{x}^{A}$ can change (Figure \ref{fig:Sketch-ghz-cat-weak-consistency},
at time $t_{4}$).

\section{Consistency of weak local realism with Bell violations}

It has been shown possible to falsify deterministic macroscopic realism
for the cat-state system described in Section V.A \cite{manushan-bell-cat-lg}.
This was demonstrated by a violation of Bell inequalities constructed
for the macroscopic spins, $|\alpha\rangle$ and $|-\alpha\rangle$.
Similarly, in Ref. \cite{manushan-bell-cat-lg}, it was shown that
wMR was not falsified by the Bell violations. This proof was expanded
in \cite{wigner-friend-macro}. Essentially, the same proof holds
to show consistency of wLR with Bell violations \cite{wigner-friend-macro}.
For the sake of completeness, we present below the proof demonstrating
consistency of wLR with Bell violations. This is relevant, since we
note that an identical proof holds to show consistency of wMR with
macroscopic Bell violations, where the spins are realised by the macroscopically
distinct states $|\uparrow\rangle_{J}\equiv|\uparrow\rangle_{J}^{\otimes N}$
and $|\downarrow\rangle_{z,J}\equiv|\downarrow\rangle_{z,J}^{\otimes N}$
and the unitary operations of the analyzer are realised by the CNOT
gates as proposed in Section V.B.

The Bell test involves the EPR-Bohm system. The Pauli spin components
defined as
\begin{eqnarray}
\hat{S}_{\theta}^{A} & = & \hat{S}_{x}^{A}\sin\theta+\hat{S}_{z}^{A}\cos\theta\nonumber \\
\hat{S}_{\phi}^{B} & = & \hat{S}_{x}^{B}\sin\phi+\hat{S}_{z}^{B}\cos\phi\label{eq:spinrot-bell}
\end{eqnarray}
can be measured by adjusting the analyzer (Stern-Gerlach apparatus
or polarizing beam splitter)  and the expectation value given by
$E(\theta,\phi)=\langle\hat{S}_{\theta}^{A}\hat{S}_{\phi}^{B}\rangle$
measured. According to the EPR-Bohm argument based on EPR's local
realism, each spin component $\hat{S}_{\theta}^{A}$ and $\hat{S}_{\phi}^{B}$
is represented by a hidden variable ($\lambda_{\theta}^{A}$ and $\lambda_{\phi}^{B}$),
because the value can be predicted with certainty by a spacelike-separated
measurement \cite{bell-1971}. This leads to the constraint $-2\leq S\leq2$
where $S=E(\theta,\phi)-E(\theta,\phi')+E(\theta',\phi)+E(\theta',\phi')$,
known as the Clauser-Horne-Shimony-Holt-(CHSH) Bell inequality, which
is violated for the Bell state (\ref{eq:bell}) \cite{bell-1971,chsh,bell-cs-review}.
The violation therefore falsifies EPR's premises based on local realism.
More generally, the violation shows failure of all local realistic
theories defined as those satisfying Bell's local realism assumptions
\cite{bell-1971,chsh}. 
\begin{figure}[t]
\begin{centering}
\includegraphics[width=1\columnwidth]{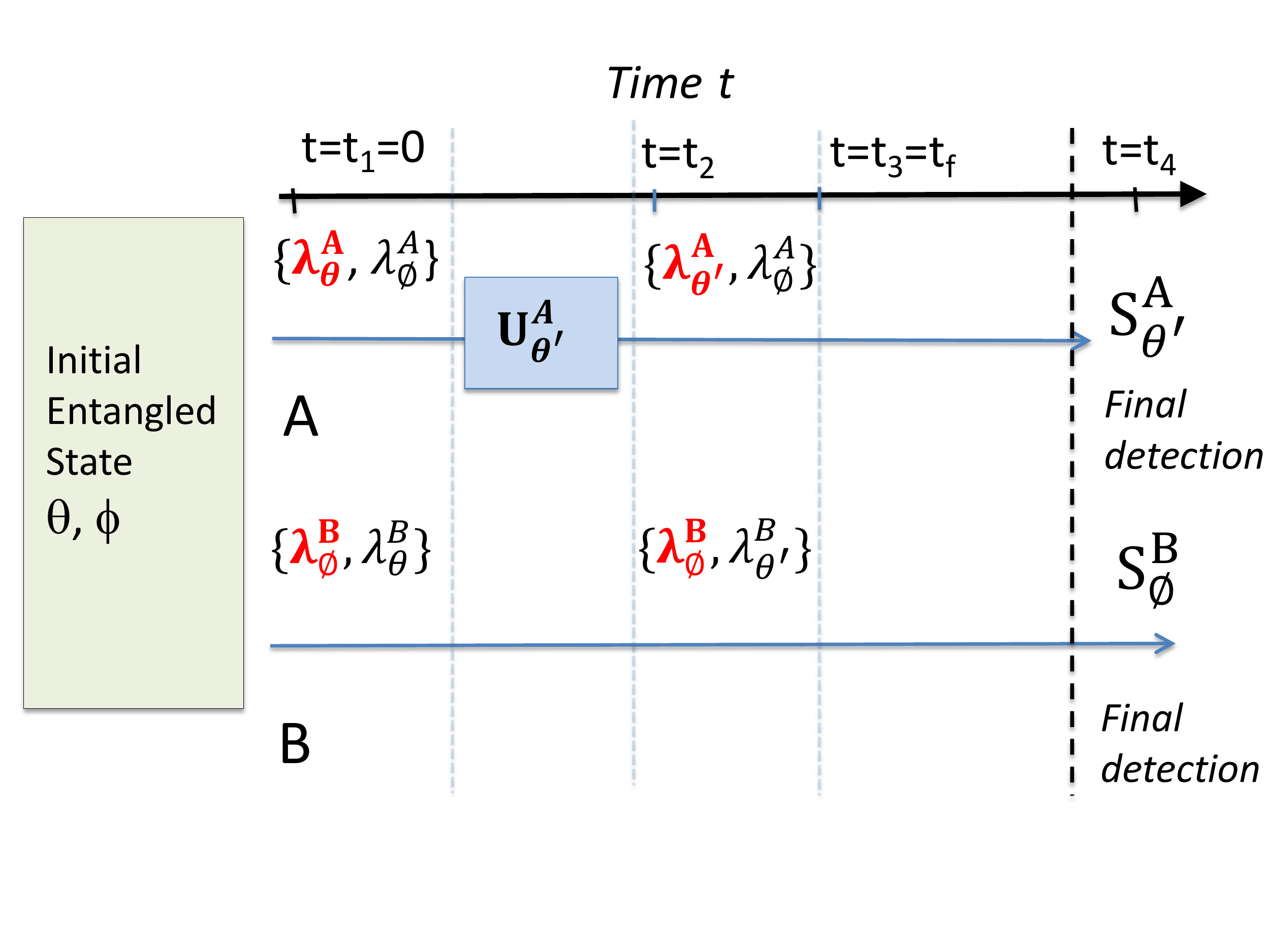}
\par\end{centering}
\begin{centering}
\includegraphics[width=1\columnwidth]{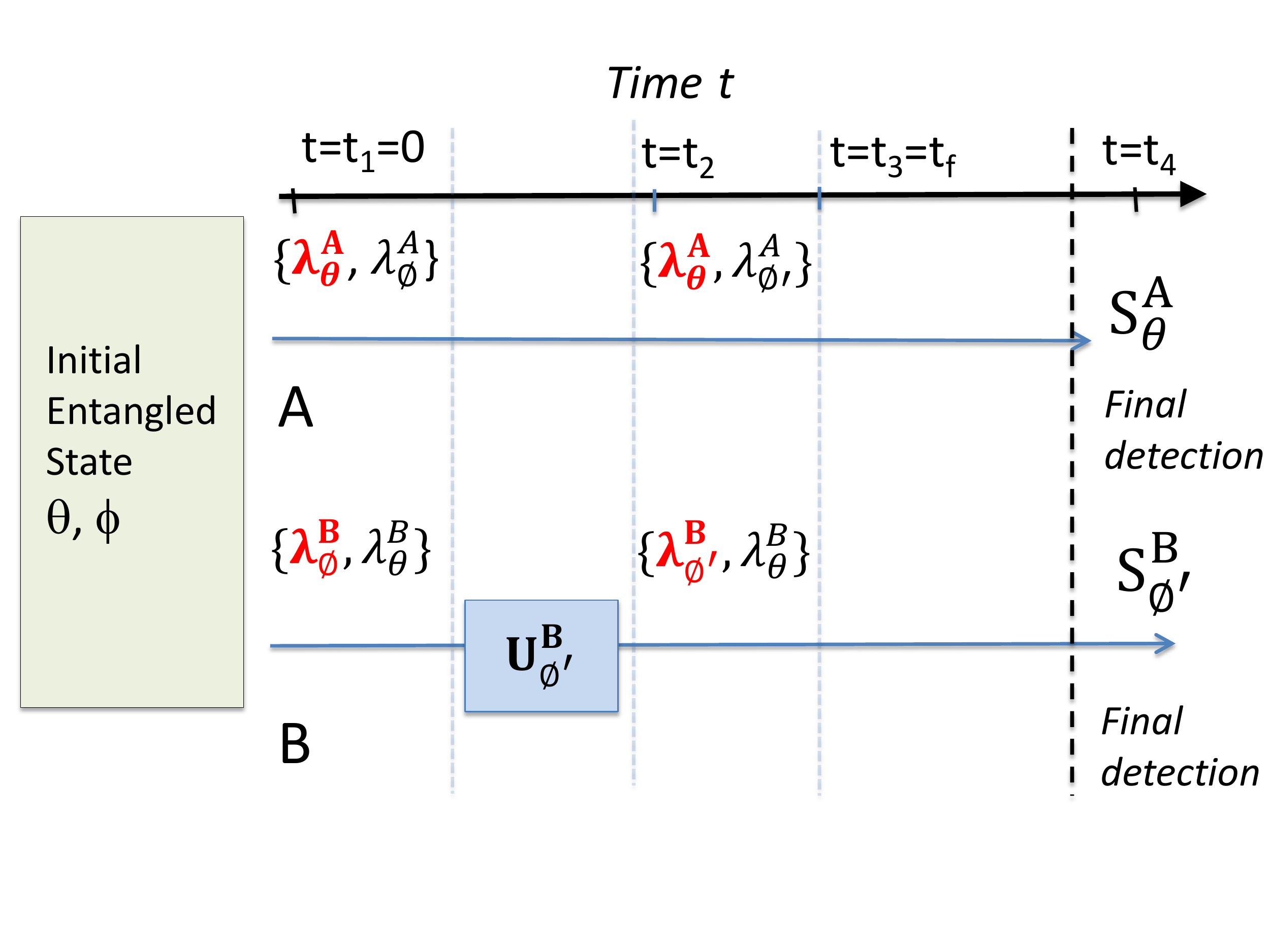}
\par\end{centering}
\caption{Tracking the hidden variables according to weak local realism (wLR)
through the dynamics of the Bell test, which shows violation of the
Bell inequality. First, the moment $\langle S_{\theta}^{A}S_{\phi}^{B}\rangle$
is measured. We take the initial time $t_{1}$ to be after the passage
through the analysers set at $\theta$ and $\phi$, so that measurement
settings $\theta$ and $\phi$ have been determined and the system
prepared for the final pointer measurements $S_{\theta}^{A}$ and
$S_{\phi}^{B}$. At each time $t_{i}$, wLR implies that certain hidden
variables $\lambda$ are valid, depending on the preparation of the
system at that time. The hidden variables are depicted in the brackets.
Those in red bold are implied by wLR Assertion 1b. Those in black
(and not bold) are implied by Assertion 1a. \label{fig:Sketch-bell-paradox-hidden}}
\end{figure}

In the Figure \ref{fig:Sketch-bell-paradox-hidden}, we track the
hidden variables that predetermine the values of the spin measurements
at each time, based on wLR. We illustrate without loss of generality
with one possible time sequence, based on the preparation at the initial
time for pointer measurements in the directions $\theta$ and $\phi$.
Assuming wLR, the values of $S_{\theta}^{A}$ and $S_{\phi}^{B}$
that are\emph{ }realised by the pointer stage of measurement (if made
at that time) are predetermined, given by $\lambda_{\theta}^{A}$
and $\lambda_{\phi}^{B}$ at the time $t_{1}$. Hence
\begin{equation}
E(\theta,\phi)=\langle\lambda_{\theta}^{A}\lambda_{\phi}^{B}\rangle.\label{eq:av1}
\end{equation}
To measure $E(\theta',\phi)$, there is a further rotation $U_{\theta'}^{A}$
at $A$. At time $t_{2}$, the state is prepared for the pointer measurements
of $S_{\theta'}^{A}$ and $S_{\phi}^{B}$. The hidden variables $\lambda_{\theta'}^{A}$
and $\lambda_{\phi}^{B}$ specify the outcomes for those pointer measurements
should they be performed. Based on the premise wLR Assertion (1b),
these variables are assigned to describe the state of the system at
the time $t_{2}$. We note that also because of the anticorrelation
evident for spins prepared in the Bell state (\ref{eq:bell}), the
wLR Assertion (1a) implies that the hidden variable $\lambda_{\phi}^{B}$
also specifies the outcome of a measurement $S_{\phi}^{A}$, if performed
on the system defined at time $t_{2}$. The prediction for wLR is
\begin{equation}
E(\theta',\phi)=\langle\lambda_{\theta'}^{A}\lambda_{\phi}^{B}\rangle.\label{eq:av2}
\end{equation}
Similarly, the measurements of $S_{\theta}^{A}$ and $S_{\phi'}^{B}$
require a further rotation $U_{\phi'}^{B}$ at $B$, after the initial
preparation at $t_{1}$, with no rotation at $A$. A variable $\lambda_{\phi'}^{B}$
is defined to give the outcome for $S_{\phi'}^{B}$, if that measurement
were to be performed at $t_{2}$ after the rotation $U_{\phi'}^{B}$.
Hence, wLR implies 
\begin{equation}
E(\theta,\phi')=\langle\lambda_{\theta}^{A}\lambda_{\phi'}^{B}\rangle.\label{eq:num}
\end{equation}

\begin{figure}[t]
\begin{centering}
\includegraphics[width=1\columnwidth]{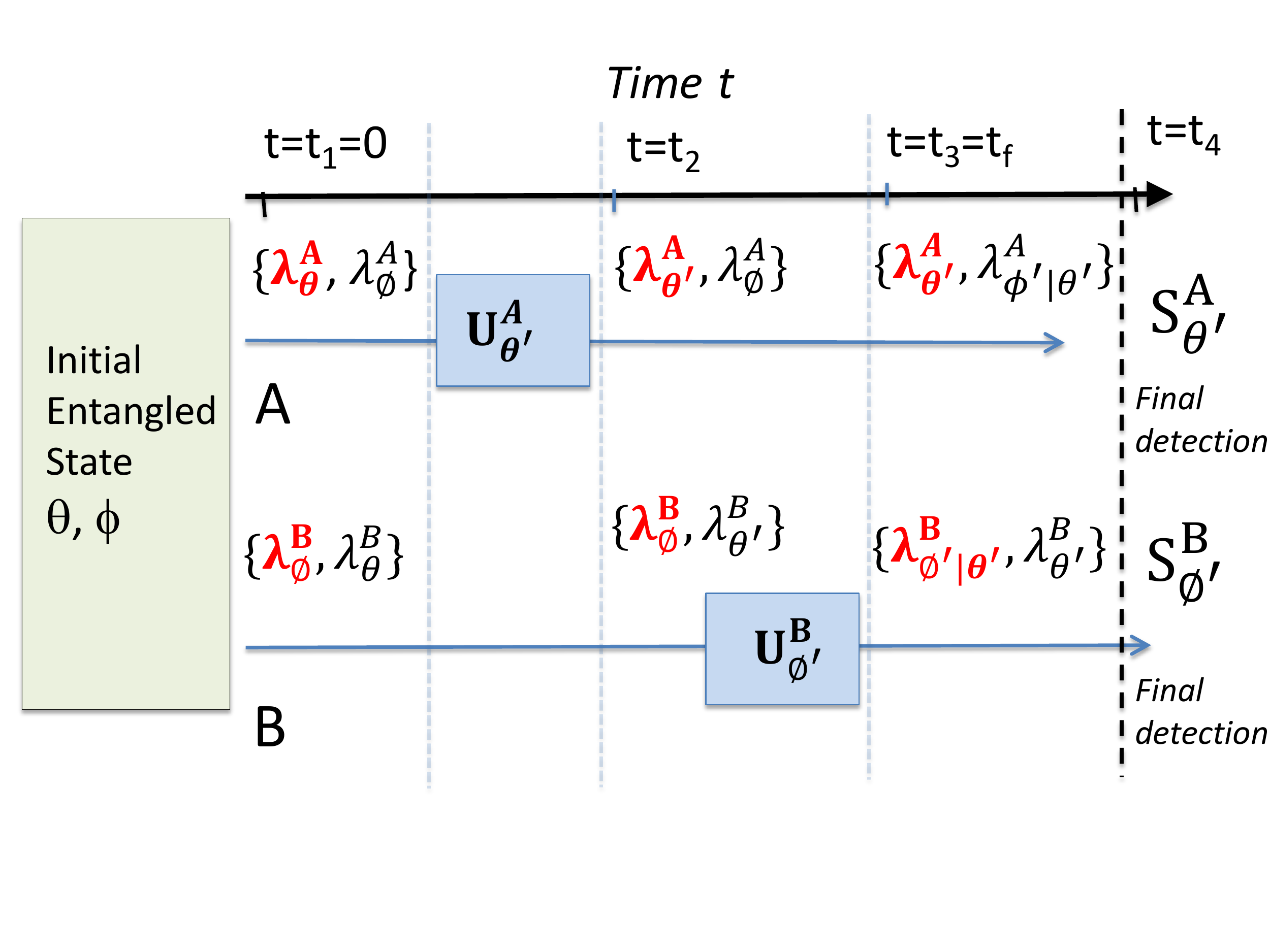}
\par\end{centering}
\caption{Tracking the hidden variables through the dynamics of the Bell test,
which shows violation of the Bell inequality. The description is as
for Figure \ref{fig:Sketch-bell-paradox-hidden}, but here the moment
$\langle S_{\theta'}^{A}S_{\phi'}^{B}\rangle$ is measured. We note
that the conditioning for $\lambda_{\phi'}$ necessitates that a rotation
$U_{\phi'}^{B}$ has also occurred at $B$, as well as $U_{\theta'}^{A}$.
The nonlocality emerges only after rotations at both sites. \label{fig:Sketch-bell-double-paradox-hidden}}
\end{figure}
The difference between Bell's local hidden variable theories and the
assertions of wLR are evident when considering measurement of $S_{\theta'}^{A}$
and $S_{\phi'}^{B}$. This measurement requires \emph{two further
rotations} after the preparation at time $t_{1}$. A possible sequence
is given in Figure \ref{fig:Sketch-bell-double-paradox-hidden}. Suppose
the rotation $U_{\theta'}^{A}$ is performed first, and at time $t_{2}$,
the hidden variables defining the state are $\lambda_{\theta'}^{A}$
and $\lambda_{\phi}^{B}$. The pointer measurements are not actually
performed, and a rotation $U_{\phi'}^{B}$ is then made at $B$. The
final state at time $t_{3}$ is given by hidden variables $\lambda_{\theta'}^{A}$
and $\lambda_{\phi'|\theta'}^{B}$. Here, we use subscripts $|\theta'$
to specify that $\lambda_{\phi'|\theta'}^{B}$ is the variable defined
for the state specified at the time $t_{3}$, conditioned on the rotation
$U_{\theta'}^{A}$ at $A$. This is necessary in the context of a
wLR model, because the premise of wLR specifies the necessity of locality
for the hidden values of the pointer measurements only. The value
$\lambda_{\theta'}^{A}$ is defined for the pointer measurement of
$S_{\theta'}^{A}$, and is independent of the choice for $\phi'$,
since this value $\lambda_{\theta'}^{A}$ is (according to wLR) not
affected by the unitary rotation $U_{\phi'}^{B}$ at the other site
$B$. However, we cannot conclude from the wLR assertions that the
value of $\lambda_{\phi'}^{B}$ defined for the measurement $S_{\phi'}^{B}$
on the state after the rotation $U_{\theta'}^{A}$ is the same as
that defined for pointer measurement $S_{\phi'}^{B}$ in Figure \ref{fig:Sketch-bell-paradox-hidden},
where there was no rotation at $A$. Hence we write
\begin{equation}
E(\theta',\phi')=\langle\lambda_{\theta'}^{A}\lambda_{\phi'|\theta'}^{B}\rangle.\label{eq:av4}
\end{equation}
We see that wLR does not imply the CHSH-Bell inequality, which is
derived based on the full Bell locality assumption that $\lambda_{\phi'}^{B}$
is independent of the value of $\theta'$ i.e. is independent of whether
the rotation $U_{\theta'}^{A}$ has been performed at $A$ or not.

It is well known that the where the values for $\lambda_{\theta}$,
$\lambda_{\phi}$, $\lambda_{\theta'}$, and $\lambda_{\phi'}$ are
either $+1$ or $-1$, and if Bell locality is assumed so that $\lambda_{\phi'|\theta}^{B}=\lambda_{\phi'}^{B}$,
then the value of $S$  is bounded by $-2$ and $2$, leading to
the CHSH-Bell inequality \cite{bell-cs-review}. However, where we
consider $\lambda_{\phi'|\theta}^{B}$ to be an independent variable,
$+1$ or $-1$, the bound for $S$ becomes the algebraic bound of
$4$. Hence, wLR does not constrain $S$ to be bounded by the Bell
inequality.

\subsection*{Nonlocality and deeper models}

The wLR and wMR premises allow for nonlocal effects, as evident by
the violation of the Bell inequality. This is the meaning of ``weak'',
that the premises do not encompass the full local realism assumptions
of EPR.

The nonlocal effect arises in the above analysis because it cannot
be assumed that the value $\lambda_{\phi'}^{B}$ is independent of
the value $\theta'$, which determines the unitary rotation at $A$.
The $\lambda_{\theta'}^{A}$ is independent of $\phi'$, because the
setting $\theta'$ is fixed (the unitary rotation $U_{\theta'}^{A}$
has occurred before $U_{\phi'}^{B}$ at $A$), but it cannot be excluded
that the $\lambda_{\phi'}^{B}$ can depend on $\theta'$ because it
occurred earlier. It seems to matter which order the unitary rotations
are taken, despite that the two rotations occur at spatially separated
locations. The predictions will not however depend on the order of
rotation. The joint distribution for values $\lambda_{\theta'}^{A}$,
$\lambda_{\phi'}^{B}$ depends on both $\theta'$ and $\phi'$, the
final settings. We note that the conditioning for $\lambda_{\phi'}$
necessitates that a rotation $U_{\phi'}^{B}$ has also occurred at
$B$, since time $t_{1}$. Rotations at \emph{both} sites are required
for the nonlocality to emerge. This feature of wMR and wLR is proved
for the GHZ set-up in Section X.  We make two comments.

First, the wLR and wMR premises remove the possibility of a strong
sort of nonlocality. In these models, the choice to measure $\phi'$
instead of $\phi$ at $B$ does not change the value of $\lambda_{\theta}^{A}$
at $A$ once the unitary rotation has occurred at $A$ to fix the
measurement setting as $\theta$ at $A$. There is hence no instantaneous
nonlocal effect. Similarly, for the system prepared in the Bell state,
the value for $S_{\phi}^{A}$ at $A$ is fixed once the rotation $U_{\phi}^{B}$
has occurred at $B$, because the value for $S_{\phi}^{A}$ can be
predicted with certainty, even when the unitary rotation $U_{\phi}^{A}$
at $A$ has not occurred. While this gives a nonlocal effect, a \emph{further}
local interaction $U_{\phi}^{A}$ is required at $A$ for the nonlocality
to be confirmed.

Second, there is motivation to examine deeper models and tests of
quantum mechanics \cite{q-contextual-1,bohm-hv,Maroney,maroney-timpson,hall-cworlds,griffiths-histories,grangier-auffeves-context,brukner-wigner,spekkens-toy-model,q-measurement,simon-q,objective-fields-entropy-1,castagnoli-2021,sabine-retro-toy,griffiths-nonlocal-not,roman-schn,philippe-grang-context-1,fr-paradox,wigner-weak,bohmian-fr,losada-wigner-friend}
for consistency with wLR and wMR. Maroney and Timpson proposed models
for macroscopic realism that allow violation of Leggett-Garg inequalities
\cite{Maroney,maroney-timpson}. In their ``supra eigenstate support
MR model'', for which there is a predetermination of the outcome
of the measurement that distinguishes between two macroscopic distinct
states, it is explained that the ``state'' of the system cannot
be an operational eigenstate, meaning it cannot be a preparable state
for the system. It is clear from their context that the system is
considered to be prepared for a pointer measurement, an example being
the observation of a ball in a box. This would give consistency with
wMR. The authors gave the de Broglie-Bohm theory as an example of
such a model. The de Broglie-Bohm theory is a nonlocal theory for
quantum mechanics \cite{bohm-hv}.

Another model of quantum mechanics that appears consistent with wMR
is the objective field theory, motivated by solutions from quantum
field theory and the Q function \foreignlanguage{australian}{\cite{q-contextual-1,q-measurement,simon-q,objective-fields-entropy-1}}.
Solutions have been given where the second stage of a measurement
is modeled dynamically as amplification of field amplitudes. Here,
there is \emph{no} direct nonlocal mechanism, but rather a retrocausality
based on future boundary conditions, which leads to hidden causal
loops \cite{castagnoli-2021}. The joint distribution for values $\lambda_{\theta'}^{A}$,
$\lambda_{\phi'}^{B}$ is shown to depend on $\theta'$ and $\phi'$,
the final settings, and Bell's local hidden variable model does not
apply. In recent work, it is reported how, for EPR and Bell correlations
based on continuous-variable measurements, the premise of wMR is upheld
\cite{q-measurement}. The premise of wMR does not allow retrocausality
at a macroscopic level because the hidden variable $\lambda$ is fixed
at the given time, being independent of any future event.

\section{Further predictions for wMR/ wLR}

We present further predictions for wMR. These provide a means to experimentally
test wMR. The predictions are identical to those of quantum mechanics.
However, wMR differs from standard quantum mechanics which does not
account for predetermined values of a measurement. The analyses apply
in identical fashion to wLR.

\subsection{Moments involving a further single rotation are consistent with local
realism}

\begin{figure}[t]
\begin{centering}
\includegraphics[width=1\columnwidth]{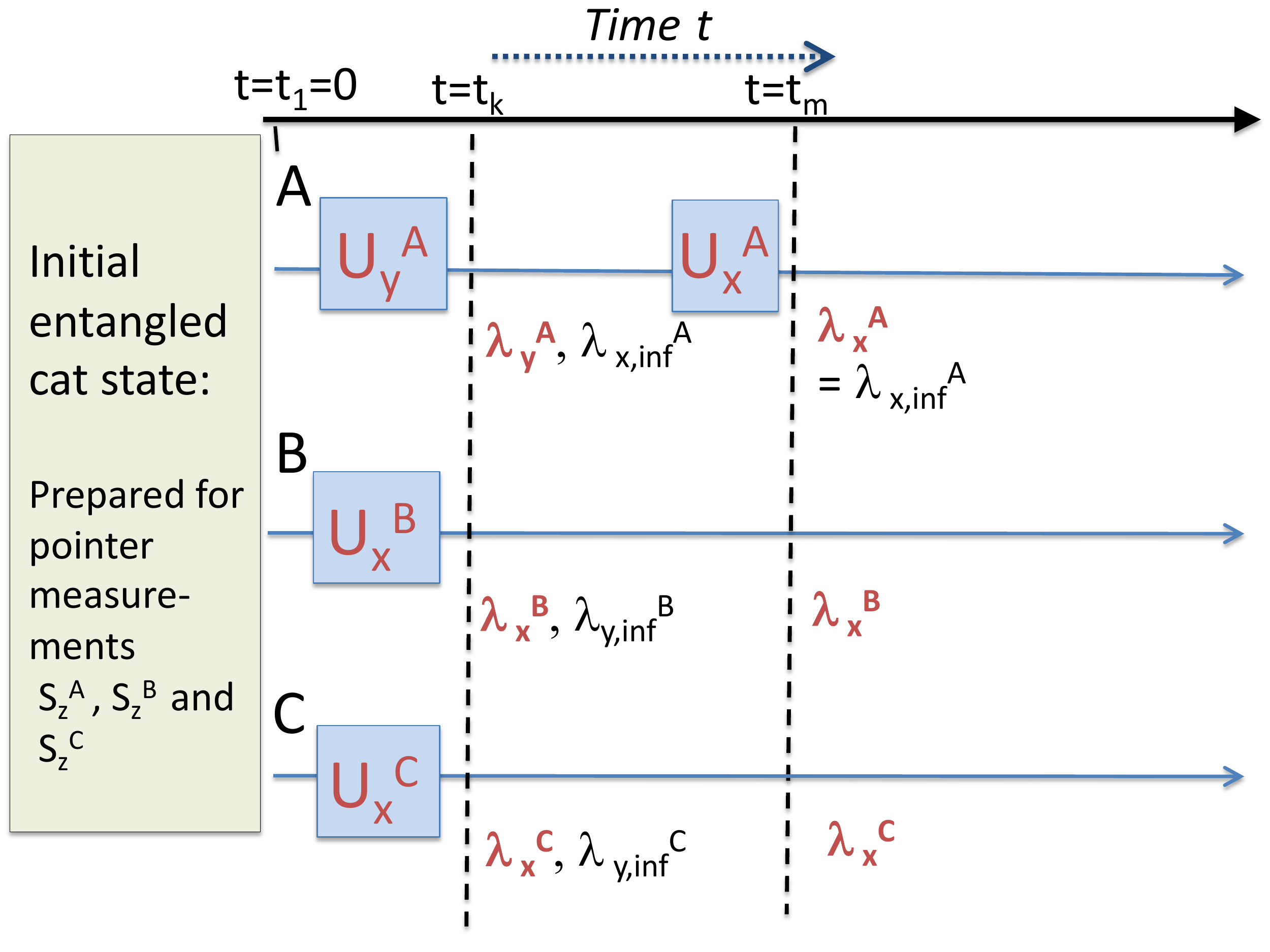}
\par\end{centering}
\caption{Predictions of weak macroscopic realism (wMR) are consistent with
local realism for the single rotation $\mathbf{U_{x}^{A}}$ after
preparation at time $t_{k}$. The notation is as for Figure \ref{fig:Sketch-ghz-cat-weak-consistency}.
The system at time $t_{k}$ is prepared such that pointer measurements
will yield outcomes for measurements of $S_{y}^{A}$, $S_{x}^{B}$
and $S_{x}^{C}$. The prediction for $S_{x}^{A}$ can be inferred
from the pointer measurements at $B$ and $C$ and hence is also predetermined
at the time $t_{k}$, according to wMR. This measurement requires
a further unitary interaction $U_{x}^{A}$. Final pointer measurements
at the time $t_{m}$ will yield the result for measurement $S_{x}^{A}S_{x}^{B}S_{x}^{C}$,
which by wMR is consistent with local realism. Quantum mechanics
also predicts consistency with local realism for the single rotation
(see text).\label{fig:Sketch-ghz-cat-weak-predictions}}
\end{figure}

\begin{figure}[t]
\begin{centering}
\includegraphics[width=1\columnwidth]{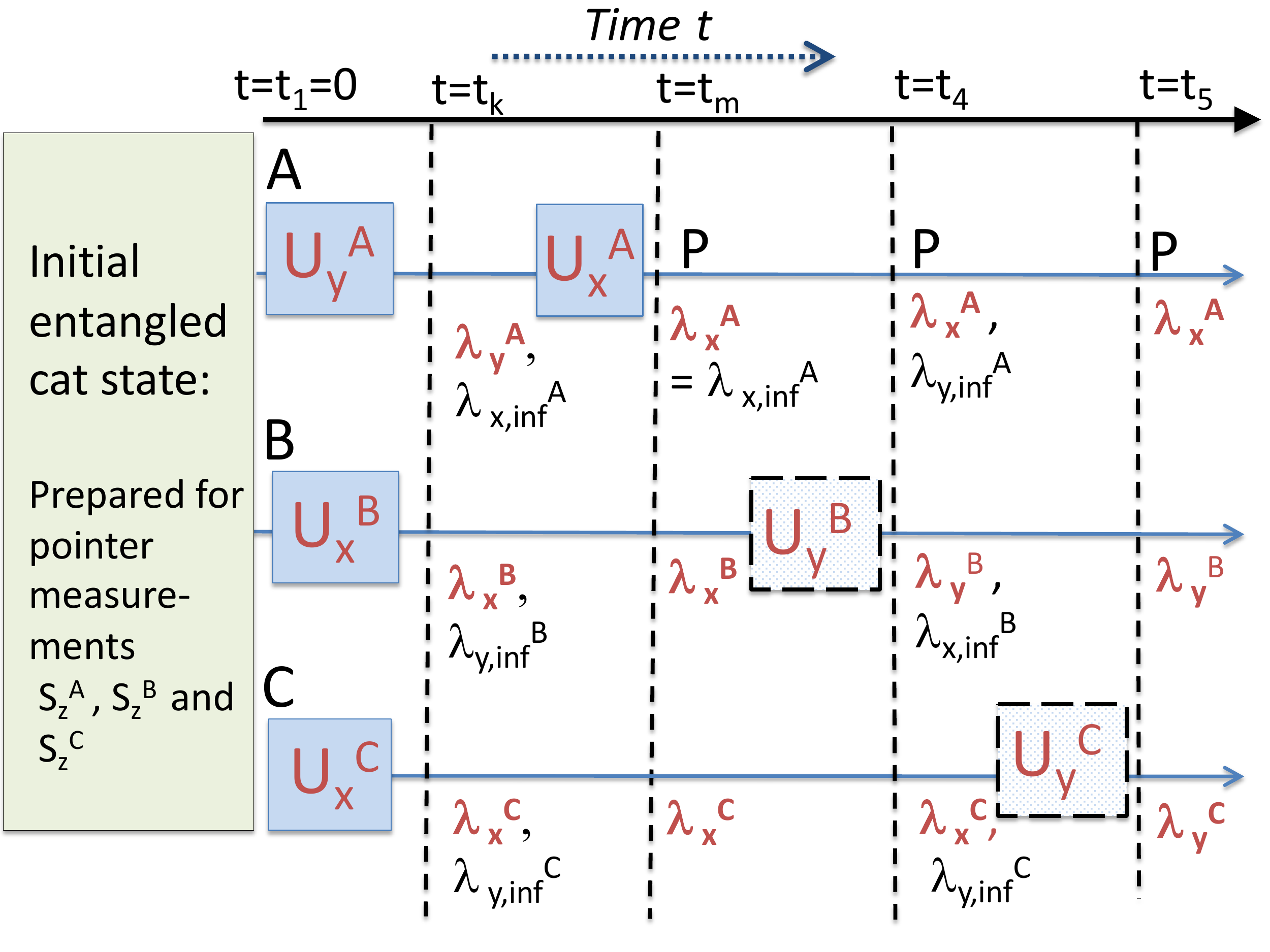}
\par\end{centering}
\caption{Predictions of weak macroscopic realism (wMR) are consistent with
those of quantum mechanics for multiple rotations. The notation is
as for Figure \ref{fig:Sketch-ghz-cat-weak-consistency}. Here, further
unitary interactions $\mathbf{U}_{y}^{B}$ and $\mathbf{U}_{y}^{C}$
prepare the system for measurement of $S_{x}^{A}S_{y}^{B}S_{y}^{C}$
at the time $t_{5}$.  However, at the time $t_{m}$, there is no
longer the pointer preparation for $S_{y}^{A}$, and the values that
were inferred for measurements $S_{y}^{B}$ and $S_{y}^{C}$ no longer
apply. The predictions lead to the GHZ paradox, but are consistent
with wMR. The premise wMR also predicts that the results for $S_{x}^{A}S_{y}^{B}S_{y}^{C}$
are independent of when the final irreversible stage of the pointer
measurement for $S_{x}^{A}$ is performed, relative to the unitary
transformations at $B$ and $C$. The different timings are indicated
by the $P$ symbol.\label{fig:Sketch-ghz-cat-weak-predictions-2}}
\end{figure}

\textbf{\emph{Prediction of wMR: }}We consider the entangled cat-state
GHZ system $|\psi\rangle$ (Eq. (\ref{eq:ghz-cat})) which is then
prepared at time $t_{k}$ for pointer measurements $S_{y}^{A}$, $S_{x}^{B}$
and $S_{x}^{C}$ at the respective sites (as in Figure \ref{fig:Sketch-ghz-cat-weak-predictions}).
The GHZ contradiction with local realism is realised by first further
changing the measurement settings, to measure $S_{x}^{A}$, $S_{x}^{B}$
and $S_{x}^{C}$, which involves one unitary rotation $\mathbf{U}_{x}^{A}$.
Also required are measurements $S_{x}^{A}$, $S_{y}^{B}$ and $S_{y}^{C}$,
which involve two further rotations, one at each site $B$ and $C$,
as well as $\mathbf{U}_{x}^{A}$ (Figure \ref{fig:Sketch-ghz-cat-weak-predictions-2}).
The prediction is that results violating local realism do not arise
from the correlations involving only one unitary $U_{x}^{A}$ after
the preparation at $t_{k}$. The violations arise from the correlations
involving the two further rotations. A similar result was proved for
Bell violations \cite{manushan-bell-cat-lg}.

\textbf{\emph{Proof:}} We denote the state prepared at the time $t_{k}$
by $|\psi\rangle_{y,x,x}$. A final pointer measurement if conducted
at time $t_{k}$ at $A$ gives the outcome for $S_{y}^{A}$. According
to wMR, at time $t_{k}$, the values for $S_{x}^{A}$, $S_{y}^{A}$,
$S_{x}^{B}$ and $S_{x}^{C}$ are each predetermined, being given
by the variables $\lambda_{inf,x}^{A}$, $\lambda_{y}^{A}$, $\lambda_{x}^{B}$
and $\lambda_{x}^{C}$. A single unitary rotation $\mathbf{U}_{x}^{A}=U_{y}^{-1}U_{x}$
at $A$ will enable the result $\lambda_{inf,x}^{A}$ to be revealed.
According to wMR, the prediction for $S_{x}^{A}S_{x}^{B}S_{x}^{C}$
is predetermined at time $t_{k}$, for the system prepared in the
state $|\psi\rangle_{y,x,x}$. There are hidden variables for each
measurement, defined for the system at $t_{k}$. Therefore, if wMR
is valid, the prediction for $S_{x}^{A}S_{x}^{B}S_{x}^{C}$, conditioned
on the initial state $|\psi\rangle_{y,x,x}$, must be entirely consistent
with local realism.$\square$

\textbf{\emph{Proof of agreement with quantum prediction:}} Here,
we prove that the prediction of quantum mechanics also shows consistency
with local realism for the set-up of just one rotation after preparation
(Figure \ref{fig:Sketch-ghz-cat-weak-predictions}). To do this, we
compare the predictions of quantum mechanics for the prepared state
$|\psi\rangle_{y,x,x}$ against those of a mixed state $\rho_{mix}^{A-BC}$.
The state $|\psi\rangle_{y,x,x}$ prepared at time $t_{k}$ in the
basis for $S_{y}^{A}$, $S_{x}^{B}$ and $S_{x}^{B}$ is
\begin{eqnarray}
|\psi\rangle_{y,x,x} & = & \frac{1}{4}\{(|\uparrow\rangle_{y}+|\downarrow\rangle_{y})(|\uparrow\rangle_{x}+|\downarrow\rangle_{x})(|\uparrow\rangle_{x}+|\downarrow\rangle_{x})\nonumber \\
 &  & +i(|\uparrow\rangle_{y}-|\downarrow\rangle_{y})(|\uparrow\rangle_{x}-|\downarrow\rangle_{x})(|\uparrow\rangle_{x}-|\downarrow\rangle_{x})\}\nonumber \\
 & = & \frac{1}{\sqrt{2}}(|\psi_{-}^{A}\rangle|\psi_{+}^{BC}\rangle+|\psi_{+}^{A}\rangle|\psi_{-}^{BC}\rangle)\label{eq:pred-1}
\end{eqnarray}
where
\begin{eqnarray*}
|\psi_{+}^{BC}\rangle & = & (|\uparrow\rangle_{x}|\uparrow\rangle_{x}+|\downarrow\rangle_{x}|\downarrow\rangle_{x})/\sqrt{2}\\
|\psi_{-}^{BC}\rangle & = & (|\downarrow\rangle_{x}|\uparrow\rangle_{x}+|\uparrow\rangle_{x}|\downarrow\rangle_{x})/\sqrt{2}
\end{eqnarray*}
and
\begin{eqnarray*}
|\psi_{+}^{A}\rangle & = & \{(1-i)|\uparrow\rangle_{y}+(1+i)|\downarrow\rangle_{y}\}/2\\
|\psi_{-}^{A}\rangle & = & \{(1+i)|\uparrow\rangle_{y}+(1-i)|\downarrow\rangle_{y}\}/2.
\end{eqnarray*}
The state $|\psi\rangle_{y,x,x}$ is a superposition involving entanglement
between the system $A$ and the system denoted $BC$, which comprises
the systems $B$ and $C$. If the unitary rotation $\mathbf{U}_{x}^{A}$
is performed at $A$, then the prediction for the pointer measurements
is $S_{x}^{A}S_{x}^{B}S_{x}^{C}=-1$. Now we compare with the system
initially prepared in the mixture 
\begin{equation}
\rho_{mix}^{A-BC}=|\psi_{-}^{A}\rangle\langle\psi_{-}^{A}|\rho_{+}^{BC}+|\psi_{+}^{A}\rangle\langle\psi_{+}^{A}|\rho_{-}^{BC}.\label{eq:mixture}
\end{equation}
Here, $\rho_{+}^{BC}=|\psi_{+}^{BC}\rangle\langle\psi_{+}^{BC}|$
and $\rho_{-}^{BC}=|\psi_{-}^{BC}\rangle\langle\psi_{-}^{BC}|$. This
mixture has no entanglement between the system $A$ and the combined
systems $B$ and $C$ i.e. it is fully separable with respect to the
bipartition that we denote by $A-BC$. If we transform to the $x$
basis at $A$, then we write 
\begin{equation}
\rho_{mix}^{A-BC}=|\downarrow\rangle_{x}\langle\downarrow|{}_{x}\rho_{+}^{BC}+|\uparrow\rangle_{x}\langle\uparrow|{}_{x}\rho_{-}^{BC}.\label{eq:mixture-x}
\end{equation}
The prediction is $S_{x}^{A}S_{x}^{B}S_{x}^{C}=-1$, which is identical
to the prediction for the system prepared in $|\psi\rangle_{y,x,x}$.
The quantum prediction for the single unitary interaction $\mathbf{U}_{x}^{A}$
on $|\psi\rangle_{y,x,x}$ is therefore consistent with local realism
$-$ since the prediction for $\rho_{mix}^{A-BC}$ is fully local
with respect to $A$, arising from a local interaction at $A$.$\square$

The GHZ test showing violation of local realism indeed requires two
further rotations, $\mathbf{U}_{y}^{B}$ and $\mathbf{U}_{y}^{C}$
at the sites $B$ and $C$, which allows measurement of $S_{x}^{A}S_{y}^{B}S_{y}^{C}$
(Figure \ref{fig:Sketch-ghz-cat-weak-predictions-2}). This is because
wMR does not (necessarily) predict for the system at time $t_{k}$,
a predetermination of the outcomes for both $S_{x}^{A}$ and $S_{y}^{B}$.
Hence, there is no contradiction between the predictions of quantum
mechanics and wMR. The hidden variables that are predicted by wMR
are tracked in the Figure \ref{fig:Sketch-ghz-cat-weak-predictions-2}.
An experiment could be performed, by comparing the observed moments
for the GHZ state with those generated by the mixed states. 

\subsection{The timing of the pointer stage of measurement}

\textbf{\emph{Prediction of wMR: }}Consider the system of Figure \ref{fig:Sketch-ghz-cat-weak-predictions-2},
prepared at time $t_{k}$ so that pointer measurements at $A$, $B$
and $C$ will give the outcomes for $S_{y}^{A}$, $S_{x}^{B}$ and
$S_{x}^{C}$. At $A$, the system is then prepared for a pointer measurement
of $S_{x}$. At $B$, a unitary rotation then prepares system $B$
for a pointer measurement of $S_{y}^{B}$, and then similarly at $C$.
If wMR is valid, then the predictions for the correlations are not
dependent on whether the final pointer stages $P$ of the measurement
for $S_{x}^{A}$ at $A$ occur before or after the unitary rotations
at $B$ and $C$. Here, the final pointer stages of the measurement
(denoted $P$ in the Figure) involve a coupling to an environment,
whereby the measurement becomes irreversible.

\textbf{\emph{Proof:}} The premise wMR asserts that the value $\lambda$
for the outcome of the pointer measurement is fixed locally for the
appropriately prepared system, provided there is no further unitary
$U$ on that system which changes the measurement setting. This prediction
agrees with that of quantum mechanics. Quantum calculations do not
distinguish the timing of the measurement stage $P$.

\section{Conclusion}

The main conclusion of this paper is that the negation of local realism,
as evidenced by a Bell or GHZ experiment, does not appear to fully
resolve the EPR paradox. In summary, we have proposed how EPR-Bohm
and GHZ experiments may be realised in mesoscopic and macroscopic
regimes, using cat states and suitable unitary interactions. These
are significant tests in a setting where all relevant measurements
are coarse-grained, distinguishing only between two macroscopically
distinct states. The macroscopic EPR-Bohm test illustrates an incompatibility
between the assumptions of deterministic macroscopic realism (dMR)
and the notion that quantum mechanics is a complete description of
physical reality. We explain it is also possible to consider the weaker
assumption, \emph{weak macroscopic realism} (wMR), and to demonstrate
a similar inconsistency with the notion that quantum mechanics is
a complete theory, using a two-spin version of the EPR-Bohm argument.
Yet, while dMR can be falsified by the GHZ experiment, the predictions
of the GHZ test agree with those of wMR. Similarly, there is no incompatibility
between wMR and violations of macroscopic Bell inequalities. In
defining wMR, it is necessary to consider that the measurement occurs
in two stages, a reversible stage establishing the measurement setting,
and an irreversible stage referred to as the pointer stage of measurement.

Similar conclusions can be drawn for the original EPR and GHZ paradoxes.
This paper motivates consideration of a weaker assumption, \emph{weak
local realism} (wLR), in the setting of the original paradox. The
EPR argument can be modified to show inconsistency between wLR and
the notion that quantum mechanics is a complete description of physical
reality. Yet, we show that the predictions of quantum mechanics for
the GHZ and Bell experiments are consistent with those of wLR. The
definitions of wMR and wLR apply to systems after the choice of measurement
basis, and hence do not conflict with the contextuality of quantum
mechanics \cite{kochen-spekker}. Our work may be seen as a supplement
to other arguments presented for the incompleteness of quantum mechanics
\cite{bell-against,bell-speakable,s-cat-1935,harrigan-spekkens,delayed-choice-cats,philippe-grang-context-1},
and may motivate a study of alternative models for quantum mechanics.

In addition to the cat-state experiments, we propose further tests
of wLR and wMR. These tests examine correlations after single unitary
rotations, and adjust the timing of the unitary interactions that
lead to the GHZ contradiction. The predictions of wMR and wLR agree
with those of quantum mechanics. The EPR and GHZ paradoxes apply
where one can predict with certainty the outcome of a measurement,
given measurements at spacelike-separated sites. Experimental factors
may prevent the realisation of predictions that are certain. The tests
can nonetheless be carried out using inequalities \cite{mermin-inequality,ardehli,belinski-klyshko,epr-rmp,bohm-test-uncertainty,epr-r2}.
Proposals for realistic tests are given in the Appendix.

The proposed experiments could be realised in the microscopic regime
using standard techniques where the unitary interactions are performed
with polarizing beam splitters. Potential macroscopic realizations
are given in Sections V and VII. The two-mode cat states involving
coherent states have been generated in cavities \cite{cat-det-map,cat-bell-wang-1},
and GHZ states generated for $N\sim20$ \cite{omran-cats}. Mesoscopic
realisations of the unitary transformations are in principle feasible
using dynamical interactions involving a nonlinear medium, or else
CNOT gates.
\begin{acknowledgments}
This work was funded through the Australian Research Council Discovery
Project scheme under Grants DP180102470 and DP190101480. The authors
thank NTT Research for their financial and technical support.
\end{acknowledgments}

\section*{Appendix}

\subsection{The unitary operation $U_{y}$ for measurement of $S_{y}$}

Consider the system $A$ originally in the eigenstate for $S_{y}:$
\begin{equation}
|\uparrow\rangle_{y}=\frac{e^{-i\pi/4}}{\sqrt{2}}(|\uparrow\rangle_{z}+i|\downarrow\rangle_{z}),\label{eq:s1}
\end{equation}
which is 
\begin{equation}
|\uparrow\rangle_{y}\equiv\frac{e^{-i\pi/4}}{\sqrt{2}}(|\alpha\rangle_{z}+i|-\alpha\rangle_{z})\label{eq:s2}
\end{equation}
in our realisation. The state after the operation $U_{y}$ is $|\alpha\rangle_{y}$,
since we see from (\ref{eq:state3-2}) that
\begin{eqnarray}
U_{y}|\uparrow\rangle_{y} & = & U_{\pi/4}^{-1}\frac{e^{-i\pi/4}}{\sqrt{2}}(|\uparrow\rangle_{z}+i|\downarrow\rangle_{z})\nonumber \\
 & = & |\alpha\rangle\label{eq:s3}
\end{eqnarray}
The pointer measurement $\hat{S}$ on this state (for large $\alpha$)
gives $+1$, corresponding to the outcome required for the eigenstate
$|\uparrow\rangle_{y}$. Similarly, consider the system prepared in
$|\downarrow\rangle_{y}$
\begin{equation}
|\downarrow\rangle_{y}=\frac{e^{-i\pi/4}}{\sqrt{2}}(|\downarrow\rangle_{z}+i|\uparrow\rangle_{z}\label{eq:s4}
\end{equation}
which is 
\begin{equation}
|\downarrow\rangle_{y}\equiv\frac{e^{-i\pi/4}}{\sqrt{2}}(|-\alpha\rangle_{z}+i|\alpha\rangle_{z})\label{eq:s5}
\end{equation}
The state after the operation $U_{y}$ is $|-\alpha\rangle_{y}$,
since from (\ref{eq:state3-2}), we see that
\begin{eqnarray}
U_{y}|\downarrow\rangle_{y} & = & |-\alpha\rangle\label{eq:s6}
\end{eqnarray}
for which the pointer measurement $X$ gives the outcome $-1$, as
required for this eigenstate. Hence, the system that is originally
in the linear superposition (\ref{eq:tr}) transforms after $U_{y}$
to
\begin{eqnarray}
U_{y}|\psi\rangle & = & d_{+}U_{y}|+\rangle_{y}+d_{-}U_{y}|-\rangle_{y}\nonumber \\
 & \rightarrow & d_{+}|\alpha\rangle+d_{-}|-\alpha\rangle\label{eq:s7}
\end{eqnarray}
As $\alpha\rightarrow\infty$, the probability of an outcome $+1$
($-1$) for the measurement $\hat{S}$ of the sign of $\hat{X}_{A}$
is $|d_{+}|^{2}$ ($|d_{-}|^{2}$) respectively, as required.

\subsection{Example of mesoscopic qubits: NOON states}

We may also consider where the macroscopic spins are two-mode number
states $|N\rangle|0\rangle$ and $|0\rangle|N\rangle$, for $N$ large.
We denote two distinct modes by symbols $+$ and $-$, and simplify
the notation so that $|N\rangle|0\rangle\equiv|N,0\rangle$ and $|0\rangle|N\rangle\equiv|0,N\rangle$.
The macroscopic qubits become $|\uparrow\rangle\rightarrow|N,0\rangle$
and $|\downarrow\rangle\rightarrow|0,N\rangle$. For the GHZ paradoxes,
we will consider three sites, labelled $A$, $B$ and $C$. There
are two modes (labelled $J+$ and $J-$) identified for each site
$J\equiv A$, $B$, $C$. The initial state would be of the form (\ref{eq:ghz-cat}).
For each site, we use the transformation \cite{macro-bell-lg}
\begin{eqnarray}
(U_{y}^{J})^{-1}|N,0\rangle_{J} & = & e^{i\varphi}(\cos\theta|N,0\rangle_{J}+i\sin\theta|0,N\rangle_{J})\nonumber \\
(U_{y}^{J})^{-1}|0,N\rangle_{J} & = & ie^{i\varphi}(\sin\theta|N,|0\rangle_{J}-i\cos\theta|0,N\rangle_{J})\nonumber \\
\label{eq:unitary-noon}
\end{eqnarray}
where $|N,0\rangle_{J}$ and $|N,0\rangle_{J}$ are the two-mode number
states at site $J$, and $\varphi$ is a phase-shift. The transformation
has been realised to an excellent approximation for $N\lesssim100$
\cite{macro-bell-lg}, using the interaction \cite{josHam-collett-steel-2-1,nonlinear-Ham-1}
\begin{equation}
H_{nl}^{J}=\kappa(\hat{a}_{J+}^{\dagger}\hat{a}_{J-}+\hat{a}_{J+}\hat{a}_{J-}^{\dagger})+g\hat{a}_{J+}^{\dagger2}\hat{a}_{J+}^{2}+g\hat{a}_{J-}^{\dagger2}\hat{a}_{J-}^{2}\label{eq:ham-noon}
\end{equation}
so that $U_{y}^{J}=e^{-iH_{nl}^{J}t/\hbar}$. Here, $\hat{a}_{J+}$,
$\hat{a}_{J-}$ are the boson destruction operators for the field
modes $J+$ and $J-$, and $\kappa$ and $g$ are the interaction
constants. The $\theta$ is a function of the interaction time $t$
and can be selected so that $\theta=\pi/4$. To realize $U_{x}^{J}$
at each site $J\equiv A,B$, $C$, we suppose the field modes $J+$
and $J-$ are spatially separated at the site $J$, so that a phase
shift $\theta_{p}$ can be applied along one arm, that of mode $J-$,
as used in the detection of NOON states \cite{noon-mitchell,noon-dowling,noon-afek,herald-noon-1-2}.
For a suitable choice of $\theta_{p}$, this induces an overall relative
phase shift between the modes, allowing realisation of the final transformation
\begin{equation}
(U_{x}^{J})^{-1}|N,0\rangle_{J}\rightarrow\cos\theta|N,0\rangle_{J}+\sin\theta|0,N\rangle_{J}.\label{eq:trans-noon}
\end{equation}

\subsection{Considerations for a realistic test of the EPR-Bohm paradox}

The EPR-Bohm paradox for the two-spin set-up of Section III.A.1 can
be signified when 
\begin{equation}
(\Delta_{inf}\hat{\sigma}_{y}^{A})^{2}+(\Delta_{inf}\hat{\sigma}_{z}^{A})^{2}<1\label{eq:steer}
\end{equation}
where $(\Delta_{inf}\hat{\sigma}_{\theta}^{A})^{2}$ is the variance
associated with the estimate inferred for the outcome of $\hat{\sigma}_{\theta}^{A}$
given a result for a measurement of $\hat{\sigma}_{\phi}^{B}$ on
system $B$. The value of $\phi$ is chosen optimally to minimize
the error \cite{epr-r2,epr-rmp}. Hence, $\phi=\theta$. A sufficient
condition that the inequality be satisfied is that 
\begin{equation}
(\Delta(\hat{\sigma}_{y}^{A}+\hat{\sigma}_{y}^{B}))^{2}+(\Delta(\hat{\sigma}_{z}^{A}+\hat{\sigma}_{z}^{B}))^{2}<1.\label{eq:steer-ineq}
\end{equation}
Then the estimate of the outcomes for $\hat{\sigma}_{\theta}^{A}$
is taken to be $-\sigma_{\theta}^{B}$, where $\sigma_{\theta}^{B}$
is the outcome of the measurement $\hat{\sigma}_{\theta}^{B}$, for
$\theta=x,y$. The bound of $1$ is half that given by Hofmann and
Takeuchi for the entanglement criterion, Eq (24) of their paper \cite{hofmann-take},
as expected for an EPR-steering inequality \cite{epr-steer,epr-rmp}.
Clearly, for the EPR-Bohm test given in Sections III and V.B, the
inequality is satisfied since there is a perfect anticorrelation between
the outcomes, implying the left-side has a value of zero. Similar
inequalities can be derived for the three-spin set-up \cite{epr-rmp}.

For the EPR-Bohm test of Section V.A, the inequality is also satisfied
in the limit of $\alpha$ large. However, for finite $\alpha$, the
states $|\alpha\rangle$ and $|-\alpha\rangle$ are not truly orthogonal.
We propose a realistic experiment as follows. Two orthogonal states
$|+\rangle$ and $|-\rangle$ are defined. The spin operators are
$\hat{\sigma}_{z}=|+\rangle\langle+|-|-\rangle\langle-|$ and $\hat{\sigma}_{y}=(|+\rangle\langle-|-|-\rangle\langle+|)/i$,
where $+$ indicates a state with an outcome for $\hat{X}$ that is
non-negative, $x\geq0$, and $-$ indicates a state with an outcome
for $\hat{X}$ that is negative, $x<0$. The notation $x\in+$ implies
$x\geq0$; the notation $x\in-$ implies $x<0$. The state being measured
is $|\psi_{Bell}\rangle$, which can be expressed as
\begin{eqnarray}
c_{1}|\alpha\rangle+c_{2}|-\alpha\rangle & = & c_{+}|+\rangle_{z}+c_{-}|-\rangle_{z}\label{eq:bell2}
\end{eqnarray}
where
\begin{equation}
|\pm\rangle=\frac{\sum_{x\in\pm}(c_{1}\langle x|\alpha\rangle+c_{2}\langle x|-\alpha\rangle)}{[\sum_{x\in\pm}|c_{1}\langle x|\alpha\rangle+c_{2}\langle x|-\alpha\rangle|^{2}]^{1/2}}|x\rangle\label{eq:statespm}
\end{equation}
and $c_{\pm}=[\sum_{x\in\pm}|c_{1}\langle x|\alpha\rangle+c_{2}\langle x|-\alpha\rangle|^{2}]^{1/2}$.
The measurement of $S_{z}$ corresponds to first determining whether
the outcome $x$ of $\hat{X}$ is non-negative or negative.
\begin{eqnarray}
P(\pm) & = & |c_{\pm}|^{2}=\sum_{x\in\pm}|c_{1}\langle x|\alpha\rangle+c_{2}\langle x|-\alpha\rangle|.^{2}\label{eq:probpm}
\end{eqnarray}
The overlap function for $\hat{x}=(\hat{a}+\hat{a}^{\dagger})/\sqrt{2}$
where $\alpha$ is real and positive is $\langle x|\alpha\rangle\sim e^{-(x-\sqrt{2}\alpha)^{2}/2}/\pi^{1/4}$
\cite{yurke-stoler-1}. Hence, 
\begin{eqnarray}
P(+) & = & \sum_{x\geq0}|c_{1}\langle x|\alpha\rangle+c_{2}\langle x|-\alpha\rangle|^{2}\nonumber \\
 & = & (\sum_{x\geq0}|c_{1}|^{2}|\langle x|\alpha\rangle|^{2}+|c_{2}|^{2}|\langle x|-\alpha\rangle|^{2}\nonumber \\
 &  & +c_{1}c_{2}^{*}\langle x|\alpha\rangle\langle-\alpha|x\rangle+c_{1}^{*}c_{2}\langle x|-\alpha\rangle\langle\alpha|x\rangle).\nonumber \\
\label{eq:proberro-1}
\end{eqnarray}
The leading term is $P(+)=\sum_{x\geq0}|c_{1}|^{2}|\langle x|\alpha\rangle|^{2}$.
The second term is an integral in the tail of the Gaussian, $e^{-(x-\sqrt{2}\alpha)^{2}}/\pi^{1/2}$,
which has mean $\mu=\sqrt{2}\alpha$ and standard deviation $\sigma=1/\sqrt{2}$.
 Taking a conservative value of $\alpha>2$, the probability of the
tail is much less than $0.03$. The third and fourth terms are damped
by a term of order $e^{-\alpha^{2}}$, and are also negligible for
$\alpha>2$. A similar result holds for $P(-)$. The error in assuming
$\langle x|-\alpha\rangle=0$ for $x\geq0$ and $\langle x|\alpha\rangle=0$
for $x<0$ introduces errors of much less than  $10$\% in the variances
of (\ref{eq:steer-ineq}). Since the uncertainty bound for the inequality
is of order $1$, the errors are negligible for $\alpha>2$.

The measurement of $S_{y}$ requires the rotation
\begin{eqnarray}
U(c_{+}|+\rangle_{z}+c_{-}|-\rangle_{z}) & = & \frac{1}{\sqrt{2}}(c_{+}-ic_{-})e^{i\pi/4}|+\rangle_{y}\nonumber \\
 &  & +\frac{1}{\sqrt{2}}(c_{+}+ic_{-})e^{-i\pi/4}|-\rangle_{y}.\nonumber \\
\label{eq:spinsimple}
\end{eqnarray}
so that the probability for an outcome $x\in+$ is 
\begin{eqnarray*}
|c_{+}-ic_{-}|^{2} & = & |c_{+}|^{2}+|c_{-}|^{2}\\
 & = & \sum_{x\in+}|c_{1}\langle x|\alpha\rangle+c_{2}\langle x|-\alpha\rangle|^{2}\\
 &  & +\sum_{x\in-}|c_{1}\langle x|\alpha\rangle+c_{2}\langle x|-\alpha\rangle|^{2}.
\end{eqnarray*}
As above, the leading terms are 
\begin{equation}
P(+)=\sum_{x\in+}|c_{1}|^{2}|\langle x|\alpha\rangle|^{2}+\sum_{x\in-}|c_{2}|^{2}|\langle x|-\alpha\rangle|^{2},\label{eq:lead}
\end{equation}
the remaining terms being negligible for $\alpha>2$. The actual measurement
is approximate, being the application of $U$ where:
\begin{eqnarray}
U(c_{1}|\alpha\rangle+c_{2}|-\alpha\rangle) & = & \frac{c_{1}}{\sqrt{2}}(e^{i\pi/4}|\alpha\rangle+e^{-i\pi/4}|-\alpha\rangle)\nonumber \\
 &  & -\frac{ic_{2}}{\sqrt{2}}(e^{i\pi/4}|\alpha\rangle+e^{-i\pi/4}|-\alpha\rangle).\nonumber \\
\label{eq:finalresult}
\end{eqnarray}
This gives 
\begin{align*}
\sum_{x\geq0}\{\frac{(c_{1}-ic_{2})}{\sqrt{2}}e^{i\pi/4}\langle x|\alpha\rangle+\frac{(c_{1}+ic_{2})}{\sqrt{2}}e^{-i\pi/4}\langle x|-\alpha\rangle\}|x\rangle\\
+\sum_{x<0}\{\frac{(c_{1}-ic_{2})}{\sqrt{2}}e^{i\pi/4}\langle x|\alpha\rangle+\frac{(c_{1}+ic_{2})}{\sqrt{2}}e^{-i\pi/4}\langle x|-\alpha\rangle\}|x\rangle.
\end{align*}
Hence, 
\begin{eqnarray}
P(+) & = & \frac{1}{2}\sum_{x\geq0}|(c_{1}-ic_{2})e^{i\pi/4}\langle x|\alpha\rangle\nonumber \\
 &  & +(c_{1}+ic_{2})e^{-i\pi/4}\langle x|-\alpha\rangle|^{2}.\label{eq:result}
\end{eqnarray}
The leading term is given by (\ref{eq:lead}), as required. Any errors
in the approximate measurement procedure lead to errors which compared
to the uncertainty bound of $1$ are negligible for $\alpha>2$.

\subsection{Considerations for a realistic test of the EPR-Bohm paradox based
on weak local realism}

The EPR-Bohm paradox based on wLR (or wMR) as described in the set-ups
of Figures \ref{fig:Sketch-bohm-test-2} and \ref{fig:Sketch-macro-bohm-paradox-1}
can be signified when 
\begin{equation}
(\Delta_{inf}\hat{\sigma}_{y}^{A})^{2}+(\Delta_{d}\hat{\sigma}_{z}^{A})^{2}<1.\label{eq:steer-1}
\end{equation}
for measurements on the system prepared at the time $t_{f}$. Here
$(\Delta_{inf}\hat{\sigma}_{y}^{A})^{2}$ is the square of the error
in the inferred value for $\hat{\sigma_{y}}^{A}$ given the result
for the measurement $\hat{\sigma}_{y}^{B}$ on system $B$. The $(\Delta_{d}\hat{\sigma}_{z}^{A})^{2}$
is the square of the error in distinguishing the spin states for the
state of system $A$ as prepared at the time $t_{f}$. The inference
variance can be measured using standard techniques, as in Appendix
C. It is also necessary to confirm that system $A$ is given quantum
mechanically as a spin $1/2$ system, which includes defining the
two spin eigenstates and demonstrating both spin measurements $\hat{\sigma}_{y}^{A}$
and $\hat{\sigma}_{z}^{A}$ (and their non-commutativity) for the
entangled system at time $t_{f}$, as well as confirming the lower
bound of the inequality (\ref{eq:uncer-spin}). The experiment of
Ref. \cite{sch-epr-exp-atom} reports simultaneous measurement along
these lines. There is no violation of the uncertainty principle for
the state of system $A$ at time $t_{f}$, because the system $A$
is in one or other spin state with equal probability, implying $(\Delta\hat{\sigma}_{z}^{A})^{2}=1$.
In the macroscopic proposals, the pseudo-spin states are the coherent
states $|\alpha\rangle$ and $|-\alpha\rangle$, or else $|\uparrow\rangle_{z}\equiv|\uparrow\rangle_{z}^{\otimes N}$
and $|\downarrow\rangle_{z}\equiv|\downarrow\rangle_{z}^{\otimes N}$.
Noise can diminish the effectiveness of the measurement $\hat{\sigma}_{z}^{A}$
, increasing $(\Delta_{d}\hat{\sigma}_{z}^{A})^{2}$. The analysis
in Appendix C indicates that the error due to the overlap of the coherent
states becomes negligible for $\alpha>2$, so that $(\Delta_{d}\hat{\sigma}_{z}^{A})^{2}\rightarrow0$.

\end{document}